\documentclass[12pt]{iopart}

\newcommand{\gf}{\textit{g}-factor } 
\newcommand{\di}[1]{\: \textnormal{#1}}
\newcommand{\tn}[1]{\textnormal{\tiny{#1}}}
\newcommand{\tN}[1]{\textnormal{#1}}
\newcommand{\CFive}{$^{12}$\tN{C}$^{5+}$} 
\newcommand{\SiThirteen}{$^{28}$\tN{Si}$^{13+}$} 
\newcommand{\SiEleven}{$^{28}$\tN{Si}$^{11+}$} 
\newcommand{\OSeven}{$^{16}$\tN{O}$^{7+}$} 
\newcommand{\Comsol}{COMSOL} 

\usepackage{iopams}  

\usepackage{graphicx}
\usepackage{float}  

\usepackage[position=t,singlelinecheck=off]{subfig}

\begin{document}

\title[The electron mass from \gf measurements on hydrogen-like carbon \CFive]{The electron mass from \gf measurements on hydrogen-like carbon \CFive} 

\author{F K\"ohler$^{1,2}$, S Sturm$^2$, A Kracke$^2$, G Werth$^3$, W Quint$^1$ and K Blaum$^2$}

\address{$^1$ GSI Helmholtzzentrum f\"ur Schwerionenforschung GmbH, Planckstra\ss{}e 1, D-64291 Darmstadt, Germany}
\address{$^2$ Max-Planck-Institut f\"ur Kernphysik, Saupfercheckweg 1, D-69117 Heidelberg, Germany}
\address{$^3$ Institut f\"ur Physik - Johannes Gutenberg-Universit\"at Mainz, Staudingerweg 7, D-55128 Mainz, Germany}
\ead{florian.koehler@mpi-hd.mpg.de}
\vspace{10pt}
\begin{indented}
\item[]Received 22 April 2015
\end{indented}

\begin{abstract}\\
The electron mass in atomic mass units has been determined with a relative uncertainty of $2.8\cdot 10^{-11}$~\cite{Strum2014_Emass}, which represents a 13-fold improvement of the 2010 CODATA value~\cite{CODATA2010}.  The underlying measurement principle combines a high-precision measurement of the Larmor-to-cyclotron frequency ratio on a single hydrogen-like carbon ion in a Penning trap with a corresponding very accurate \gf calculation. Here, we present the measurement results in detail, including a comprehensive discussion of the systematic shifts and their uncertainties. A special focus is set on the various sources of phase jitters, which are essential for the understanding of the applied line-shape model for the \gf resonance.
\end{abstract}

%
\vspace{2pc}

\section{Introduction}
\begin{figure}[htb]
\centering
\includegraphics[width=0.8\textwidth]{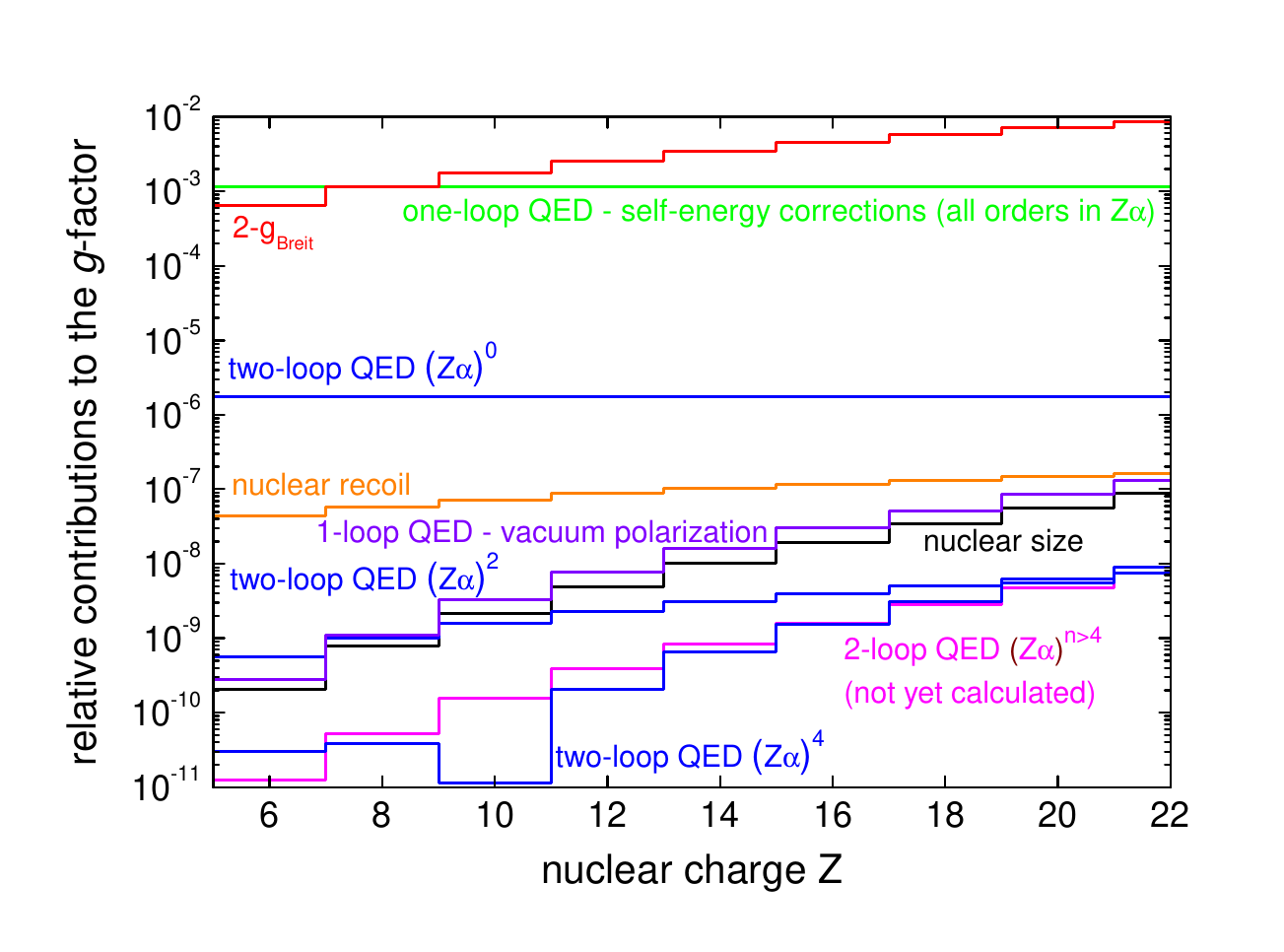}
\caption{\label{fig:gfactorTheory}Relative \gf contributions for hydrogen-like ions without nuclear spin as a function of the nuclear charge, $Z$ \cite{TheoryZoltan}.}
\end{figure}

As the lightest charged particle of the Standard Model of physics the electron is one of the few fundamental building blocks of our visible universe. The atomic structure of matter and its dynamics is profoundly determined by the intrinsic properties of the electron, most notably its mass, $m_e$. The precise knowledge of this quantity is essential for tests of the underlying theory, the quantum electrodynamics (QED).\\  
The presently most common approach in high-precision mass spectrometry of charged particles is the measurement of the cyclotron frequency ratio of the particle of interest $\nu_c^\tn{int}=\frac{1}{2\pi}\frac{q_\tn{int}}{m_\tn{int}}B$ and a reference particle $\nu_c^\tn{ref}=\frac{1}{2\pi}\frac{q_\tn{ref}}{m_\tn{ref}}B$ in the same, homogeneous magnetic field, $B$:
\begin{equation}
\label{eqn:directApproach}
m_\tn{int}=\frac{q_\tn{int}}{q_\tn{ref}}\frac{\nu_c^\tn{ref}}{\nu_c^\tn{int}}m_\tn{ref},
\end{equation}
where $q_\tn{int}$ and $q_\tn{ref}$ are the charge states of the particle of interest and of the reference particle, respectively, and $m_\tn{ref}$ is the mass of the reference particle, which should have a smaller relative uncertainty than the mass of interest $m_\tn{int}$. In high-precision mass spectrometry it is a particular challenge to measure both cyclotron frequencies in the same magnetic field either by simultaneous measurements in multiple traps or subsequent measurements in the same trap, requiring a very stable magnetic field. \\
For the determination of the electron mass this direct approach has been applied for the first time by G. G\"artner and E. Klempt in 1978~\cite{Gaertner1978}. 
\begin{figure}[htb]
\centering
\includegraphics[width=0.8\textwidth]{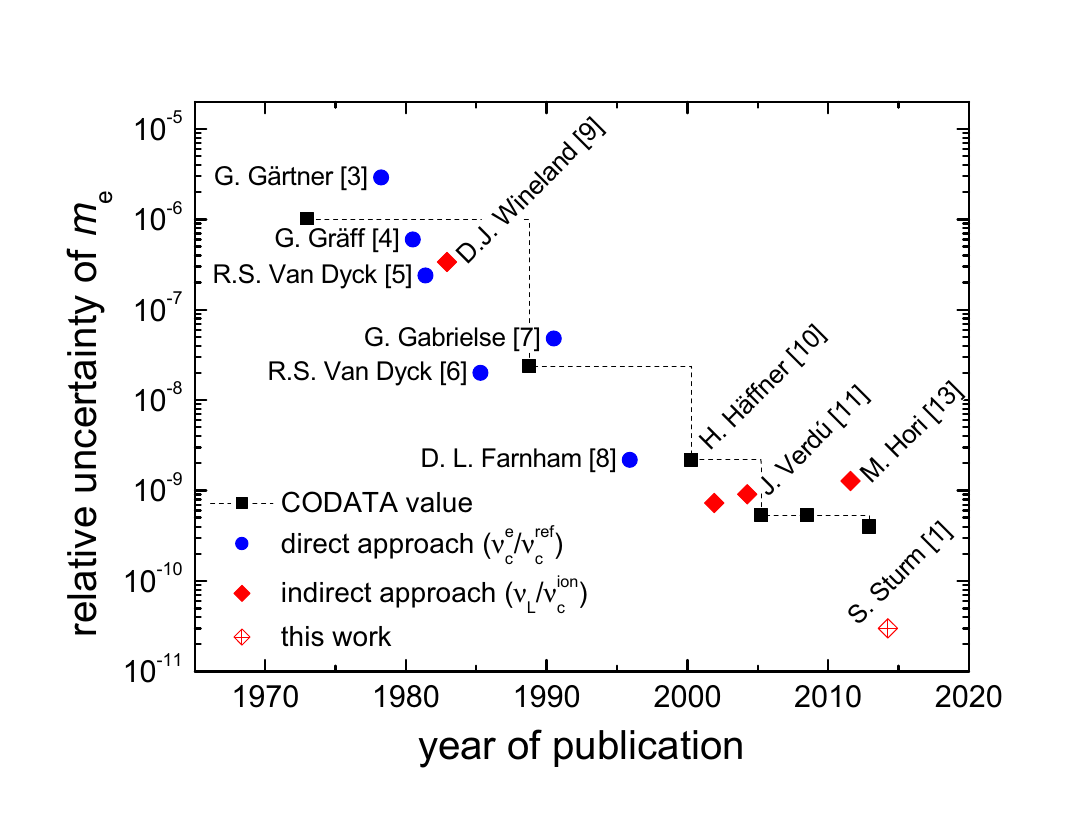}
\caption{\label{fig:EmassHistory}A historical overview of the relative uncertainties of electron mass measurements in the last 40 years. Blue markers: direct measurements of the cyclotron frequency ratio $\nu_c^\tn{ref}/\nu_c^\tn{e}$; red markers: indirect measurements of the Larmor-to-cyclotron frequency ratio $\nu_L/\nu_c^\tn{ion}$; black markers: literature values by CODATA. Next to the markers the first authors and the references are listed.}
\end{figure}They loaded alternately clouds of electrons and protons into a Penning trap and measured in a destructive manner the particle loss after resonant excitation of the cyclotron motion. In the following 17 years the size of the particle clouds was reduced and the frequency detection techniques were improved~\cite{Graeff1980,VanDyckJr1981,VanDyckJr1985,Gabrielse1990}. The most recent direct electron mass determination was performed by D. L. Farnham and others in 1995~\cite{Farnham1995}, measuring the cyclotron frequencies of a single carbon nucleus and a small cloud of electrons alternately in a Penning trap by using the induced image currents for detection. This measurement principle is limited by the large relativistic shift and the corresponding large uncertainty of the electron cyclotron frequency due to the smallness of the electron mass. In figure~\ref{fig:EmassHistory} the relative uncertainties of various electron mass measurements of the last 40 years are summarized.\\
In 1982 D. Wineland and others~\cite{Wineland1983} developed an alternative approach for precision mass spectrometry in a Penning trap, which circumvents the image charge detection for the determination of the cyclotron frequency. They measured on the one hand the cyclotron frequency of a small cloud of lithium-like $^9\tN{Be}^+$ ions, using a laser induced fluorescence technique, and on the other hand a hyperfine transition of these ions via an rf-optical double-resonance technique. For the determination of the electron mass they had to rely on theoretical predictions for the bound-electron \textit{g}-factor.\\ 
In the beginning of this millennium, G. Werth, H.-J. Kluge and W. Quint presented an improved and advanced variant of this indirect approach to circumvent the determination of the cyclotron frequency of the free electron and the corresponding undesired large relativistic shifts \cite{Haeffner2000,Haeffner2001}. For the first time a single hydrogen-like ion, stored in a double Penning-trap setup, has been used for the determination of the electron mass. Here, the Zeeman splitting of the bound electron provides a relation between the electron mass and the spin precession frequency, the Larmor frequency, $\nu_L$:
\begin{equation}
\label{eqn:ZeemanSplitting}
\Delta E= h \nu_L = h \frac{g}{4 \pi}\frac{e}{m_e}B,
\end{equation}
where $h$ is the Planck constant, $g$ is the bound-electron \gf and $B$ is the field strength of the homogeneous magnetic field the ion is stored in. The magnetic field is determined by the measurement of the cyclotron frequency, $\nu_c,$ of the ion itself. The measurement of the frequency ratio, Larmor-to-cyclotron frequency, enables an indirect determination of the electron mass in the case $g$ is known from QED theory:
\begin{equation}
\label{eqn:IndirectApproach}
m_e= m_\tn{ion}\frac{g}{2}\frac{e}{q_\tn{ion}}\frac{\nu_c}{\nu_L}\equiv  m_\tn{ion}\frac{g}{2}\frac{e}{q_\tn{ion}}\Gamma.
\end{equation}
In 2000 and 2003 this ratio has been measured using \CFive ~by H. H\"affner~\cite{Haeffner2000} and using \OSeven ~by J. Verd\'u~\cite{Verdu2004}\footnote{One further indirect measurement has been performed by M. Hori and colleagues in 2011, using two-photon laser spectroscopy of antiprotonic helium, a bound-state including a helium nucleus, an antiproton and an electron. Relying on matter-antimatter symmetry, $m_p=m_{\overline{p}},$ they derived a value for the electron mass~\cite{Hori_2011}.}.\\ 
The extractions of the electron mass from these measurements rely on theoretical calculations of the bound-electron's \textit{g}-factor. Various physical effects contribute to its value, e.g. the Breit term ($g_\tn{Breit}=2/3(1+2\sqrt{1-(Z\alpha)^2})$) \cite{Breit1928} from the solution of the Dirac equation for an electron bound in a field of a point-like charge $Ze$, nuclear properties (finite size and mass) and QED loop corrections, which can be calculated in the framework of bound-state quantum-electrodynamics (BS-QED)~\cite{Volotka2013}. Their relative contributions to the \gf of a hydrogen-like ion are shown in figure~\ref{fig:gfactorTheory} as a function of the nuclear charge, $Z$.
The one-loop QED contributions (6 Feynman diagrams) have been numerically calculated in all orders of $(Z \alpha$). For small $Z$ the two-loop QED contributions (50 Feynman diagrams) are calculated more precisely in an expansion of $(Z \alpha),$ which has been evaluated up to the fourth order. The most accurate BS-QED calculations exist for ions with a small nuclear charge, due to the scaling of up-to-now uncalculated two-loop Feynman diagrams of the order $(Z \alpha)^{5}$ or higher, which presently represent the dominant theoretical uncertainty. In 2011 BS-QED calculations have been tested with so far unprecedented precision by the \gf measurement on a single hydrogen-like silicon ion,\SiThirteen, ($Z=14$)~\cite{Sturm2011_hydrogenlikeSi}. In that case theory and experiment are in agreement at the  $8.5\cdot10^{-10}$ level.\\ 
Similar to H. H\"affner, we measure the Larmor-to-cyclotron frequency ratio of a single hydrogen-like carbon ion, \CFive, ($Z$=6), using new and significantly improved frequency detection techniques. 
State-of-the-art calculations~\cite{Strum2014_Emass} provide the electronic \gf of \CFive ~with a relative uncertainty of $2.4\cdot10^{-12}$. 
Since the unified atomic mass unit is defined by $1/12$ of the mass of a single $^{12}$C atom, the uncertainty of the mass of \CFive ~($\delta m/m=1.5\cdot10^{-12}$) is primarily given by the uncertainty of the binding energies of the five missing electrons.\\ 
In the following the measurement of the frequency ratio $\Gamma \equiv \nu_L / \nu_c$ is outlined in detail. 

\section{Measurement techniques for bound-electron \textit{g}-factors}
\begin{figure}[htb]
\centering
\includegraphics[width=1.0\textwidth]{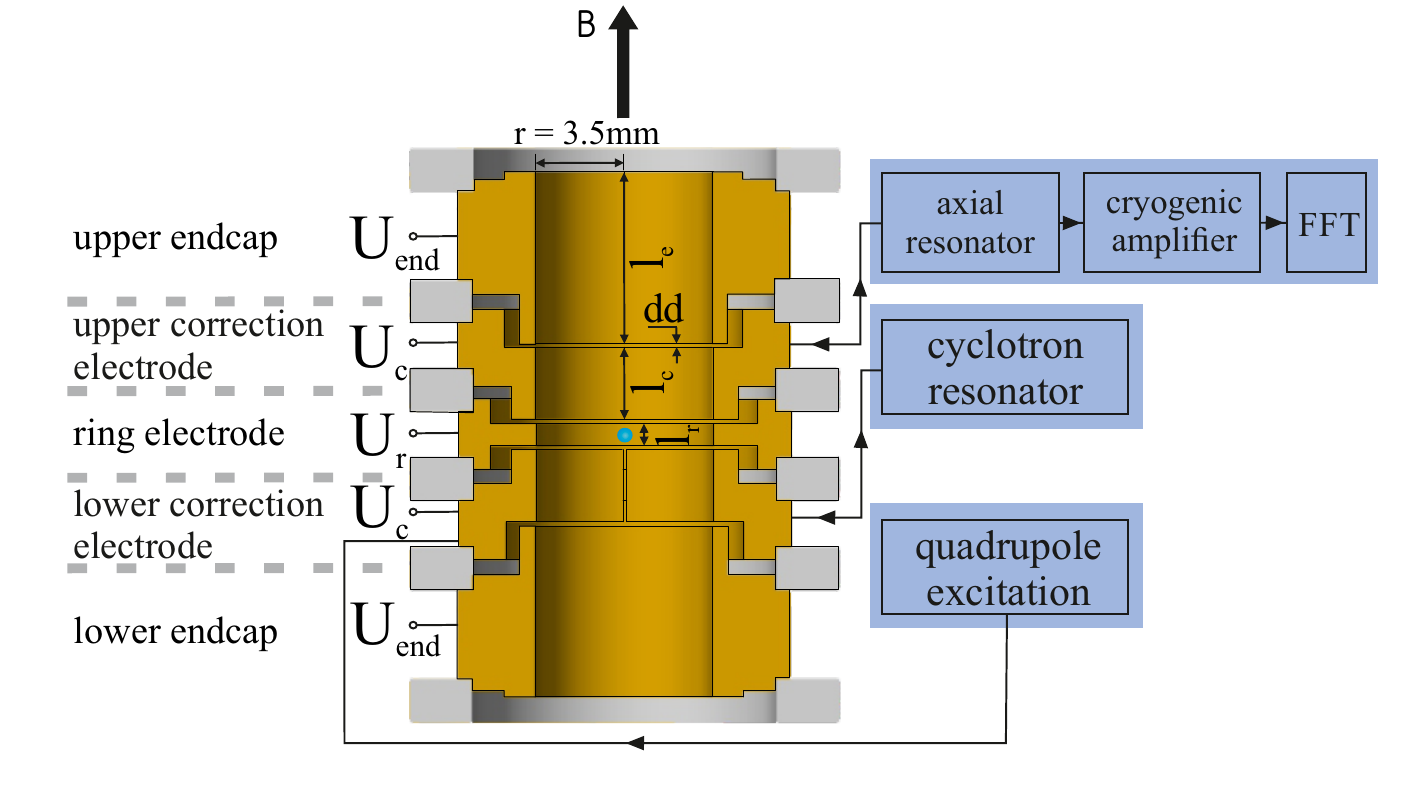}
\caption{\label{fig:ExpSetup}Cross section of the precision trap (PT), which is a 5-fold segmented cylindrical Penning trap. The induced image charges on the \textit{upper correction electrode} are detected to measure the axial frequency, $\nu_z$. The \textit{lower correction electrode} is split, so that a quadrupole excitation can be applied and the induced image current at the cyclotron frequency can be detected with a dedicated tank circuit.}
\end{figure}

Our experimental setup has already been used for the previous high-precision \gf measurements of \SiThirteen~\cite{Sturm2011_hydrogenlikeSi} and \SiEleven~\cite{Wagner2013_lithiumlikeSi} and is described in detail in~\cite{Schabinger2012}. Additional implementations for the present experiment comprise a solenoidal self-shielding coil~\cite{Wagner_PhD} for an improved temporal stability of the magnetic field and a more compact realization of the phase-sensitive technique (PnA) to measure the ion's modified cyclotron frequency~\cite{Sturm2011_PnA}.\\
The core of the experimental setup is a stack of five electrodes (inner radius: $r=3.5\di{mm}$), made of gold-plated oxygen-free high conductivity (OFHC) copper, which form a cylindrical Penning trap. 
Figure~\ref{fig:ExpSetup} shows a cross section of the precision trap. 
Proper voltages applied to the five electrodes - the central ring electrode, two correction electrodes and two endcaps - generate an electrostatic quadrupole potential, which on the one hand confines the single ion in axial direction to a harmonic \textit{axial eigenmotion} (amplitude: $z$, frequency: $\nu_z$), but on the other hand repels the ion in radial direction. A homogeneous magnetic field of nominally $3.76\di{T},$ directed along the symmetry axis of the trap, forces the ion on a cyclotron orbit in the radial plane of the trap. It is generated by a superconducting, shimmed solenoid with an inner bore diameter of $12.4\di{cm}.$ 
The $\vec{E} \times \vec{B}$ field-configuration in the radial plane splits the \textit{free cyclotron motion} into two radial harmonic eigenmotions: the fast \textit{modified cyclotron motion} ($r_+,\nu_+$) and the slow \textit{magnetron motion} ($r_-,\nu_-$). For the electric and magnetic field parameters, as chosen in our experiment, the \CFive ~ion has the following three eigenfrequencies:
\begin{eqnarray}
\nu_z &=\frac{1}{2 \pi}\sqrt{\frac{q_\tn{ion} \; U_r\; C_2}{m_\tn{ion} \;d_\tn{char}^2}}\approx 670\;964 \di{Hz} \label{eq:axialFreq},\\
\nu_+ & = \frac{1}{2}\left(\nu_c+\sqrt{\nu_c^2-2\nu_z^2}\right) \approx 24\;081\;134 \di{Hz} \label{eq:modCyclFreq},\\
\nu_- & = \frac{1}{2}\left(\nu_c-\sqrt{\nu_c^2-2\nu_z^2}\right) \approx 9347\di{Hz} \label{eq:magnFreq}.
\end{eqnarray}
$U_r=-7.634\di{V}$ is the voltage at the ring electrode, $d_\tn{char}=3.083\di{mm}$ is a characteristic trap dimension\footnote{We define the characteristic trap length as: $d_\tn{char} \equiv \sqrt{0.5(z_0^2+r^2/2)},$ where $z_0 \equiv l_r /2+dd+l_c+dd$ ($l_r=0.92\di{mm},$ $l_c=2.85\di{mm}$ and $dd=0.14\di{mm}$), see figure~\ref{fig:ExpSetup}.} and $C_2$ is a dimensionless parameter from the harmonic term of the axial expansion of the trapping potential:
\begin{equation}
\label{eqn:Potential:zaxis}
V(z)=\frac{1}{2} U_\tn{r} \sum_{i=0,2,4} C_i \frac{z^i}{d_{char}^i}.
\end{equation}
In this expansion odd terms can be neglected due to rotational and mirror symmetry of the trap. Frequency shifts, given by higher-order imperfections of the electric potential, will be discussed in section~\ref{sec:elImper}. The voltages applied to the electrodes are generated by an ultra-stable voltage source (UM-1-14~\cite{UM114}). During the relevant measurement time of $8\di{min.},$ the relative voltage fluctuations are smaller than $6\cdot10^{-8}$.\\
The free cyclotron frequency of the ion is determined from the three eigenfrequencies via the \textit{invariance theorem}~\cite{Brown1986_Geonium}: 
\begin{equation}
\label{eqn:InvarTheorem}
\nu_c=\sqrt{\nu_+^2+\nu_z^2+\nu_-^2}\approx 24\;090\;481 \di{Hz},
\end{equation}
where eigenfrequency shifts due to trap imperfections, as ellipticity of the trapping potential and a tilt of the magnetic field direction with respect to the trap axis, cancel.

\subsection{Eigenfrequency measurement techniques}
\begin{figure}
\subfloat[]{\includegraphics[width=0.49\textwidth]{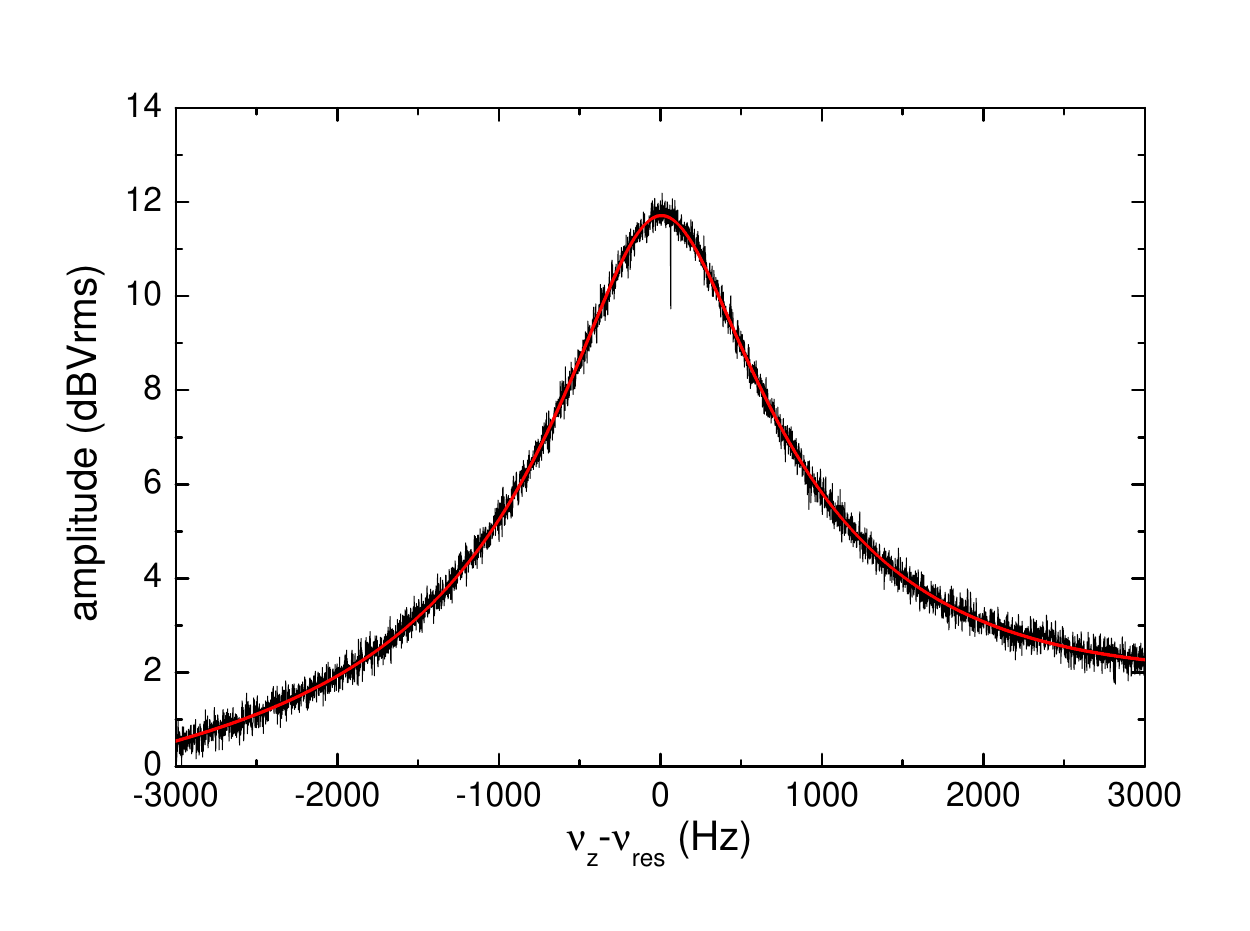}}\hfill
\subfloat[]{\includegraphics[width=0.49\textwidth]{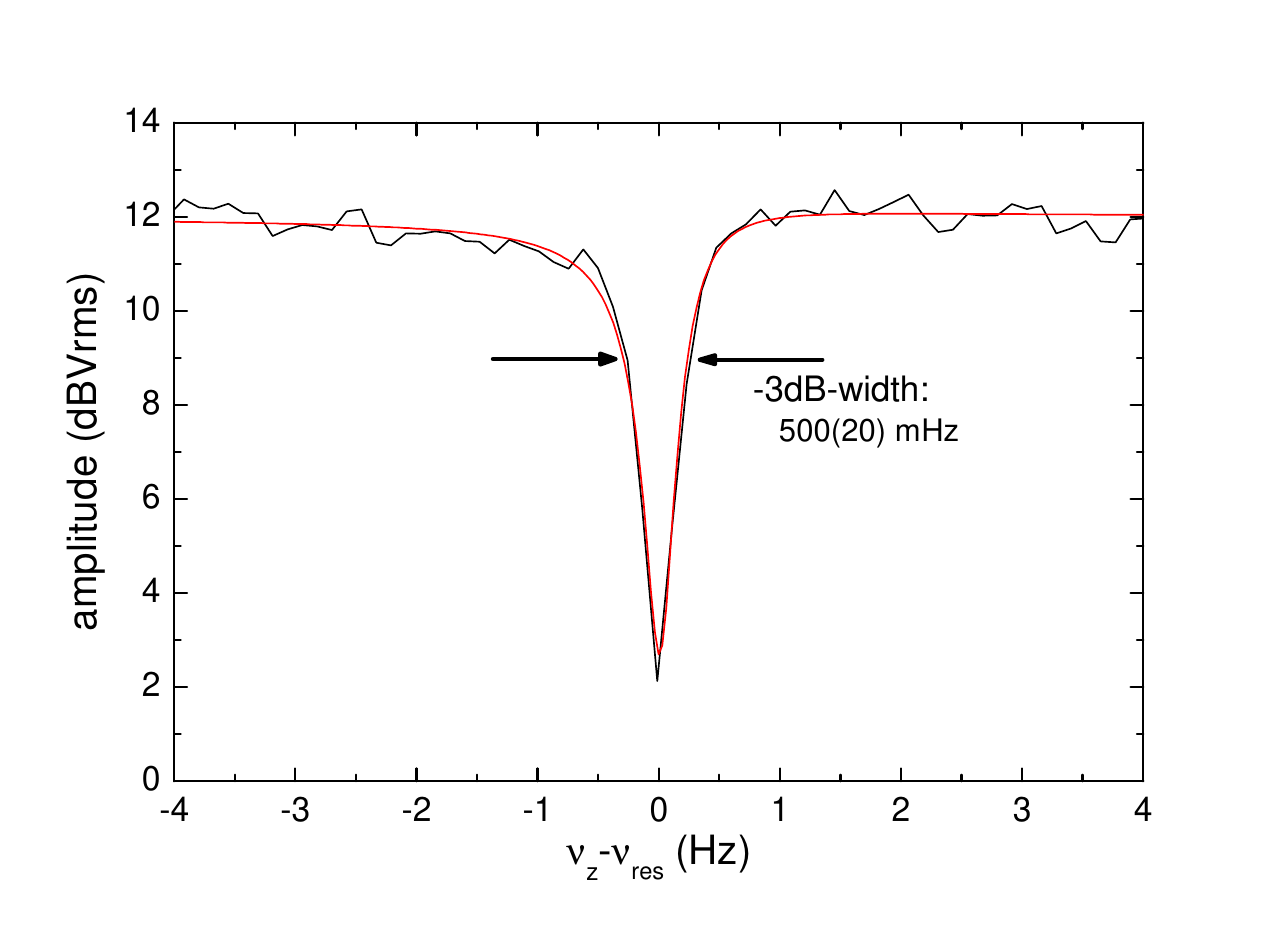}}
\caption{\label{fig:AxialResonatorAndFit}Fourier spectrum of the axial resonator in the precision trap. (a) Thermal noise spectrum of the resonator with an unresolved dip-signal and the least-squares fitted line-shape model (red). (b) High resolution dip-signal of a single axially thermalized \CFive ~ion (black) and the line-shape fit (red).}
\end{figure}

The oscillating ion induces image currents on the surfaces of the trap electrodes. At typical kinetic energies of the axial mode, corresponding to an ion temperature of about $4\di{K}$ temperature, the induced current on the undivided upper correction electrode is in the order of a few femtoamperes. This current is transformed into a measurable voltage drop by a large parallel resistance of $6.8\di{M}\Omega,$ which is generated by a superconducting solenoidal resonator with a quality factor of $Q=670$, see figure~\ref{fig:AxialResonatorAndFit}(a). Already in the cryogenic section the axial signal is amplified by a dedicated ultra-low noise amplifier~\cite{Sturm_PhD}. 
If the tank circuit and the axial oscillation frequency of the ion are kept in resonance, the axial motion thermalizes with the resonator and the axial energy decreases exponentially with a cooling time constant of $\tau \equiv m_\tn{ion}\; D_\tn{eff} \; q_\tn{ion}^{-2} \; \tN{Re}(Z(\nu_Z))^{-1} =331(14)\di{ms},$ where $D_\tn{eff}=7.38\di{mm}$ is the \textit{effective electrode distance}\footnote{The effective electrode distance is defined as $D_\tn{eff} \equiv U_\tn{el}/\frac{\partial U_\tn{el}}{\partial z}|_{z=0,r=0}$, where $U_\tn{el}$ is an electric potential applied to the electrode the axial tank circuit is connected to. Furthermore, the calculation of $D_\tn{eff}$ requires an electric field configuration, where all other electrodes are set to zero voltage and the ion is absent, see also~\cite{RamoTheo}.} and $\tN{Re}(Z(\nu_Z))$ is the real part of the resonator impedance. After the thermalization process the ion's axial oscillation amplitude is Boltzmann distributed. Measuring the axial amplitude fluctuations, we determine the ion's axial temperature to $T_z=5.44(22)\di{K}$. The equivalent electronic circuit of the oscillating ion is a series resonance circuit which shortens the thermal noise (Johnson noise) of the resonator, $n_J(\nu_z)=\sqrt{4 k_B \;  T_z \; \tN{Re}\left[Z(\nu_z)\right]},$ leading to a minimum in the Fourier-transform of the noise spectrum. 
The axial frequency can be extracted from a fit to the well-known line-shape~\cite{Sturm_PhD} of the so-called \textit{dip}-signal, see figure~\ref{fig:AxialResonatorAndFit}(b).
The averaged $\tN{-}3\tN{dB}$-dip width, $\Delta(\nu_z)_\tn{-3dB}=(2 \pi \tau)^{-1}=480(20)\di{mHz}$ scales linearly with the number of thermalized \CFive ~ions.\\
For the determination of the radial oscillation frequencies, we couple these motions to the axial motion via quadrupole rf-sideband excitations, e.g. for the modified cyclotron motion at the lower sideband $\nu_\tn{rf} \approx \nu_+-\nu_z.$ Here, the axial motion turns into a dressed state and the single dip-signal splits up into two dips (\textit{double-dip}), 
\begin{figure}[htb]
\centering
\includegraphics[width=0.8\textwidth]{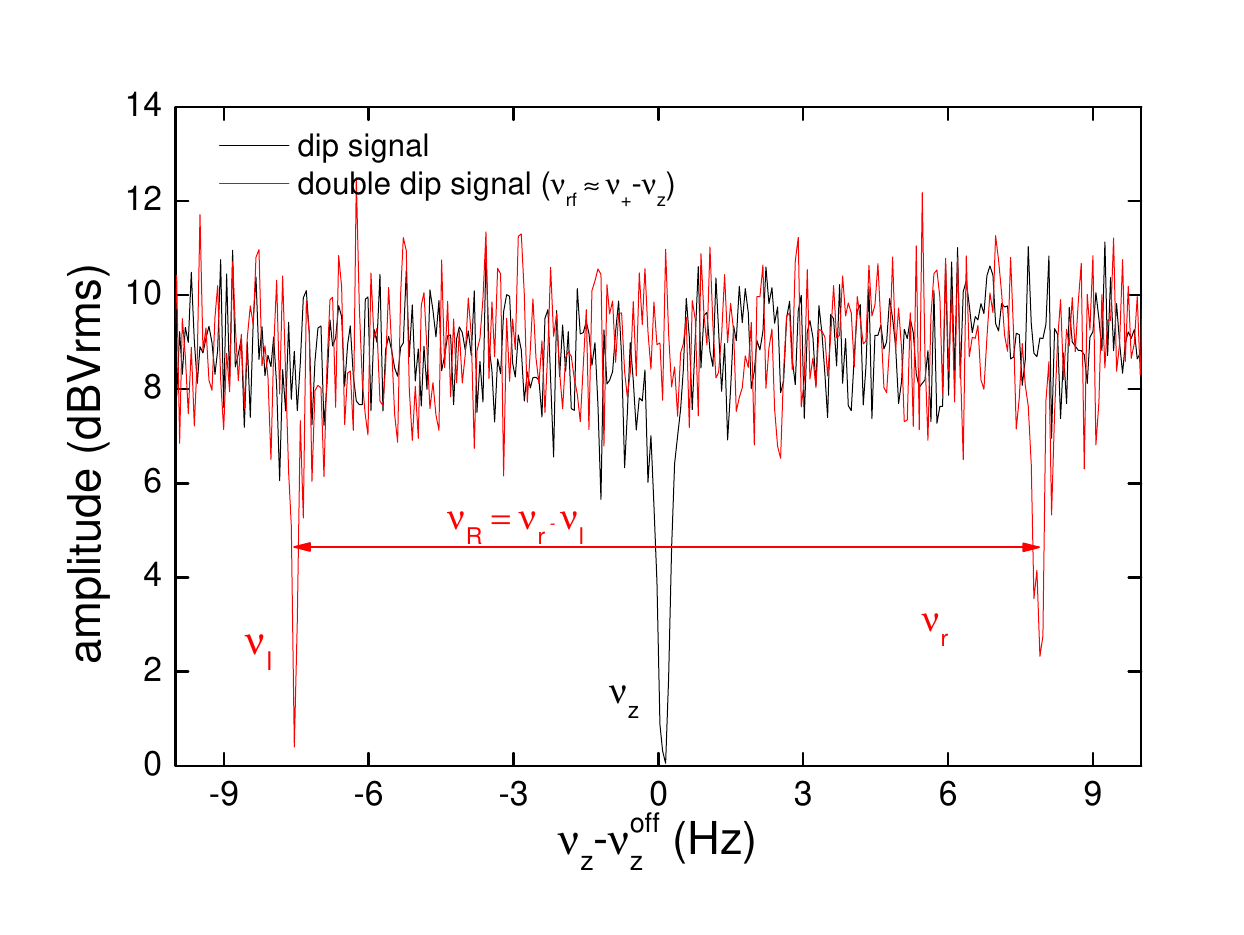}
\caption{\label{fig:DDSignal}Axial frequency spectra showing either a single axial dip-signal (black, uncoupled) or a double-dip signal (red, coupled), if the modified cyclotron mode is coupled to the axial mode via rf-sideband coupling. For details see text.}
\end{figure}with a left ($\nu_l$) and a right ($\nu_r$) dip, see figure~\ref{fig:DDSignal}. Finally the modified cyclotron frequency can be derived from the relation:
\begin{equation}
\label{eqn:vpDD}
\nu_+=\nu_\tn{rf}-\nu_z+\nu_l+\nu_r.
\end{equation}
During the coupling the energies of the modified cyclotron mode and the axial mode oscillate with the Rabi frequency $\nu_R=\nu_r-\nu_l,$ which is proportional to the amplitude of the sideband coupling. For our chosen coupling amplitude, we have  $\nu_R \approx 16\di{Hz}.$ Due to the continuous exchange of action between the axial and cyclotron mode and the energy dissipation by the cryogenic axial resonator, the modified cyclotron motion is also thermalized via the axial resonator. In thermal equilibrium we arrive at a temperature of the modified cyclotron motion of $T_+=\nu_+/\nu_z\cdot T_z=35.9\cdot T_z.$\\
In a similar manner the magnetron mode can be coupled to the axial motion for the frequency determination and cooling (reduction of the magnetron radius). Here the upper sideband $\nu_\tn{rf}=\nu_z+\nu_-$ is applied, since the magnetron energy consists of almost exclusively potential energy and decreases with increasing magnetron radius, resulting in a negative magnetron temperature. The magnetron frequency is determined via:
\begin{equation}
\label{eqn:vmDD}
\nu_-=\nu_\tn{rf}+\nu_z-\nu_l-\nu_r.
\end{equation}
 Using only the axial resonator for cooling, the energies of the eigenmotions can be calculated as follows: $\bar{E}_z=\frac{1}{2} m \omega_z^2 \bar{z^2}=k_B T_z,$ $\bar{E}_+ \approx \frac{1}{2} m \omega_+^2 \bar{r_+^2}=k_B T_+$ and $\bar{E}_- \approx -\frac{1}{4} m \omega_z^2 \bar{r_-^2}=k_B T_-.$ Measured temperatures, amplitudes and energies of a single \CFive ~ion are listed in table~\ref{tab:radii}.\\
During the phase-sensitive measurements of the modified cyclotron frequency, see section~\ref{sec:PnA}, the temperature is even further reduced, below the ambient temperatures of $4.2\di{K},$ by applying negative feedback to the axial resonator~\cite{Sturm_PhD}.
\begin{table}
\centering
\caption{\label{tab:radii}Temperatures, amplitudes and energies of the three eigenmotions for a single \CFive ~ion, thermalized by the axial resonator in the PT.}
\footnotesize
\begin{tabular}{c | lll}
\br
mode & axial & mod. cycl. & magnetron\\
\mr
temperature (K) & 5.44(22) & 195(8) & -0.076(4)\\ 
amplitude ($\mu$m) &  21(1) & 3.4(1) & 3.4(1)\\
energy (meV) & 0.47(2) & 16.8(7) &  -0.0065(3)\\
\br
\end{tabular}
\end{table}
\normalsize

\subsubsection{Phase-sensitive detection technique of $\nu_+$ - The PnA method}\label{sec:PnA}
\begin{figure}[htb!]
\centering
\includegraphics[width=1\textwidth]{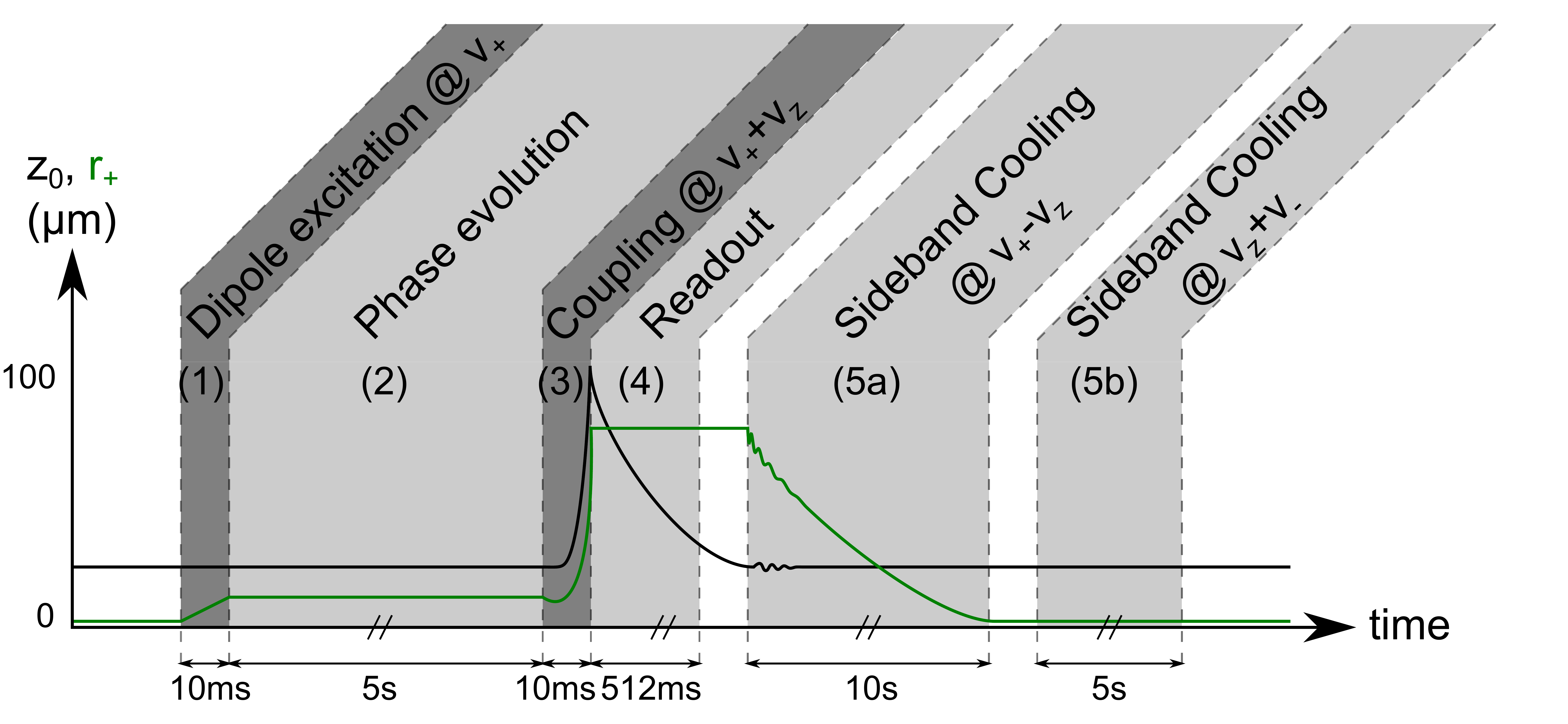}
\caption{\label{fig:PnAScheme}Schematic of the phase-sensitive detection technique PnA (Pulse and Amplify) ~\cite{Sturm2011_PnA}. The five-step PnA-cycle is illustrated by the progressions of the axial amplitude (black line) and the modified cyclotron radius (green line). For details see text.}
\end{figure}

~\\
Since $\nu_+$ represents the largest contribution to the free cyclotron frequency, see equations (\ref{eq:axialFreq}-\ref{eq:magnFreq}) and (\ref{eqn:InvarTheorem}), it has to be measured with highest precision. 
A dedicated phase-sensitive technique, the PnA (Pulse and Amplify) method~\cite{Sturm2011_PnA}, has been used for a fast measurement of the modified cyclotron frequency at low energies. In the following we introduce the five-step \textit{PnA-cycle}, see figure~\ref{fig:PnAScheme}, which is crucial for the understanding of the later discussion on measurement uncertainties.\\
(1) Starting with a thermalized ion, the modified cyclotron motion is excited by a short ($10\di{ms}$) radial dipole radio frequency pulse ($\nu_\tn{rf} \approx \nu_+$) with a fixed starting phase, resulting in a well-defined modified cyclotron amplitude and phase. (2) In the following, the modified cyclotron mode is completely decoupled from any detection system and the phase of this mode freely evolves for a phase evolution time $T_\tn{evol}$ of typically $5\di{s}$ at a motional radius of $r_+^\tn{evol}=13\; \mu\tN{m}$ (or larger). (3) Subsequently a second, quadrupole pulse at the expected upper sideband, $\nu_\tn{rf}=\nu_+ + \nu_z,$  parametrically excites the axial and the modified cyclotron mode for about $10 \di{ms}$. Since $r_+^\tn{evol}>z \sqrt{\nu_z/\nu_+},$ the phase information of the modified cyclotron mode is transferred into the phase of the axial mode during this second pulse. Simultaneously both amplitudes increase exponentially. Detailed information, e.g. the differential equations and the corresponding solutions, are given in~\cite{Sturm2011_PnA}. (4) Right after the second pulse the excited ion is detected as a peak signal above the thermal noise spectrum of the axial resonator. The readout time is typically $512\di{ms},$ which corresponds roughly to two times the cooling time constant. From the complex amplitudes of the Fourier transform we determine the axial phase of the ion, which contains the modified cyclotron phase information. (5) Finally both radial modes are thermalized again via rf-sideband coupling to the axial resonator; for the modified cyclotron mode the coupling time is $10\di{s}$ and for the magnetron mode the coupling time is $2\di{s}.$ In general, even lower modified cyclotron temperatures in the Kelvin regime could be reached with a cyclotron resonator. However, technical problems with our cyclotron resonator introduced in table~\ref{tab:CyclotronReso} lead to a small quality factor, which finally result in unacceptable cooling time constants of several  minutes.\\
\begin{figure}[htb]
\centering
\includegraphics[width=0.8\textwidth]{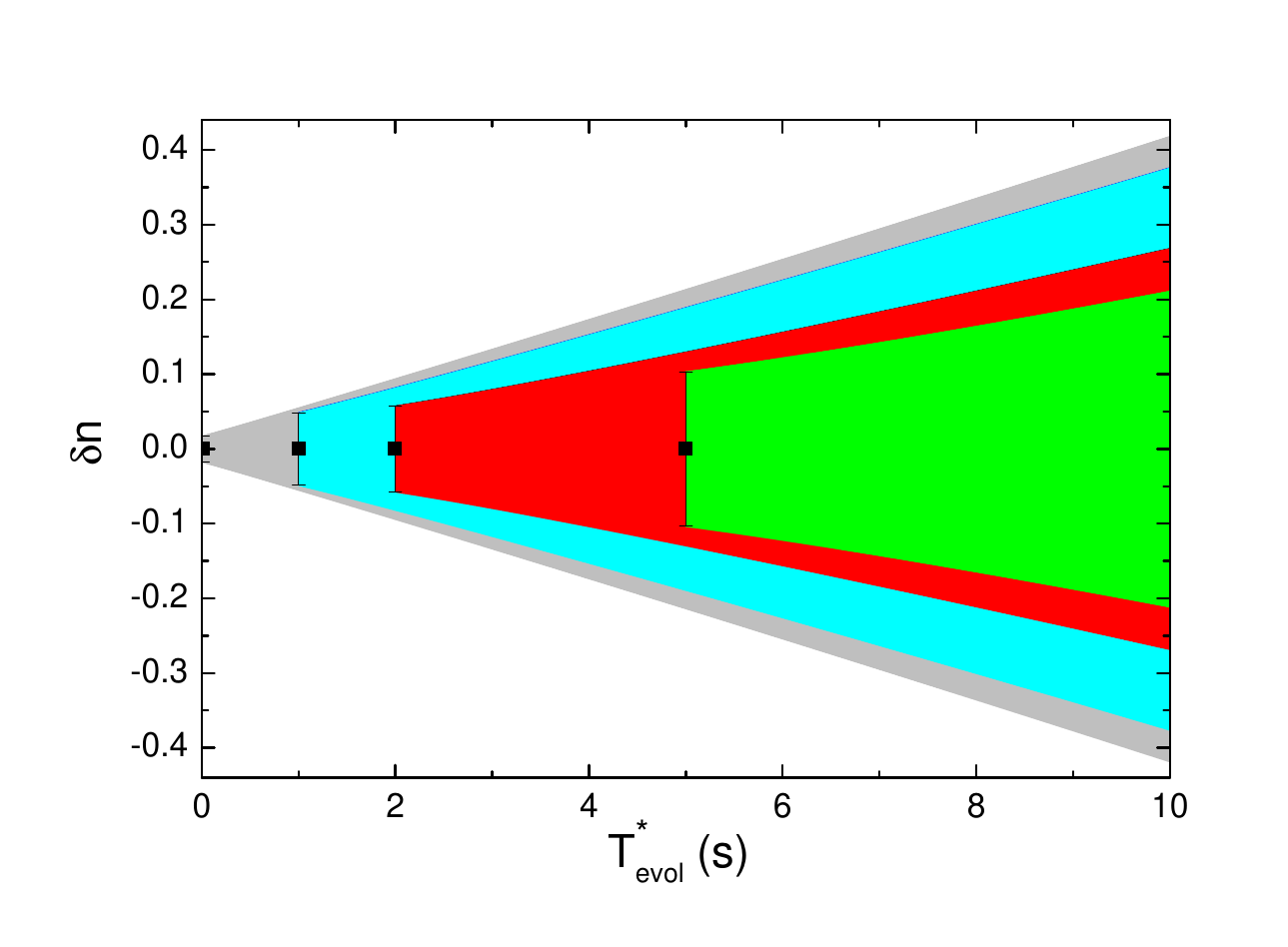}
\caption{\label{fig:PhaseUnwrapping}The principle of phase unwrapping. The colored error-bands illustrate the uncertainty of the phase evolution number $\delta n$ as a function of the phase evolution time. The gray band is determined by the phase uncertainty of the six averaged $10\di{ms}$-PnA-cycles and the frequency uncertainty from a previous $3\di{min}$ double-dip measurement. This extrapolated $\delta n$ is subsequently reduced by PnA phase measurements with phase evolution times of 1 s (cyan), 2 s (red) and 5 s (green). For further details see text.}
\end{figure}A phase-sensitive frequency determination requires at least two phase measurements at two different point of times. One phase, the starting phase, is measured with very short evolution times of $T_\tn{evol}^\tn{start}=10\di{ms}.$ At these short time scales the phase jitter due to magnetic field fluctuations is negligible compared to other contributions. We repeat these $10~\tN{ms}$-PnA-cycles six times, to average and reduce the uncertainty of the starting phase. A detailed discussion of the different sources of phase jitter is given in section~\ref{sec:jittersources}. As the frequency uncertainty scales inversely with the measurement time, $T_\tn{evol}^*= T_\tn{evol}^\tn{final}-T_\tn{evol}^\tn{start}:$ \begin{equation}
\label{eqn:PnAResol}
\frac{\delta \nu_+}{\nu_+}=\frac{\delta \varphi}{360^\circ \cdot \nu_+ \cdot T_\tn{evol}^*},
\end{equation}
the phase evolution time $T_\tn{evol}^\tn{final}$ of the second phase measurement should be as long as possible.\\
Since phases can only be determined modulo a factor of $360^\circ,$ an appropriate phase unwrapping is required to determine the absolute phase: $\varphi_\tn{abs}=n\cdot 360^\circ+\varphi_\tn{meas},$ where $n$ is an integer. With the knowledge of $\nu_+$ from a previous \textit{double-dip} measurement and the starting phase from the $10\di{ms}$-PnA cycles, the phase evolution number, $n,$ can be predicted for longer phase evolution times: $n(T_\tn{evol})=\left |\left[\varphi_\tn{10 ms}+360^\circ\cdot\nu_+^\tn{DD}\cdot (T_\tn{evol}-10\tN{ms})\right]/360^\circ \right|$. For a proper phase unwrapping the predicted uncertainty of the phase evolution number should be well below $1$, see figure~\ref{fig:PhaseUnwrapping}. In this plot the gray band illustrates the extrapolated phase evolution number uncertainty in dependence of the phase evolution time, given by six  $10\di{ms}$-PnA-cycles and the frequency uncertainty from a previous $3\di{min}$ double-dip measurement. This estimated phase uncertainty is subsequently reduced by PnA cycles with phase evolution times of 1 s (cyan), 2 s (red) and 5 s (green). For example at a phase evolution time of 5 s the uncertainty $\delta n$ based on the uncertainties of $\nu_+^\tn{DD}$ and $\varphi_\tn{10 ms}$ (gray band) is roughly halved by the 2 s PnA measurement (red band). The maximal phase evolution time and thus the frequency uncertainty are limited by magnetic field fluctuations, which cause a significant probability for phase unwrapping errors ($3\cdot \delta \varphi > 180^\circ$) at phase evolution times larger than $8\di{s}$. In figure~\ref{fig:PhaseUnwrapping} magnetic field fluctuations have been also considered, resulting in a slight non-linear increase of the phase evolution number uncertainty.\\ 
The complete pulse sequence of a single PnA cycle: ''dipole pulse - waiting time - couple/amplification pulse'' is programmed as an arbitrary waveform of a two-channel function generator, Agilent 33522A, connected to one half of the split correction electrode, indicated by ''quadrupole excitation'' in figure~\ref{fig:ExpSetup}. This arbitrary function generator and the spectrum analyzer (SR1) are triggered by a pulse delay generator (BNC 555); all three devices are locked to a rubidium atomic clock (SRS FS725).\\
In comparison to the double-dip measurement of $\nu_+,$ which requires a long averaging time of about $3 \di{min}$ and reaches a relative Allan deviation of $\delta \nu_+ / \nu_+=2\cdot 10^{-9},$ 
the PnA method only requires a measurement time of $5\tN{-}10~\di{s}$  (neglecting the \textit{preparation times}: the time for the double-dip measurement, the six $10\di{ms}$-PnA cycles, the phase unwrapping PnA-cycles (1 s and 2 s) and the time for the cooling of the modes between these cycles) for a relative Allan deviation of $\delta \nu_+ / \nu_+\approx 5\cdot 10^{-10},$ see also figure~\ref{fig:MagneticFieldJitter}. 
In contrast to the double-dip technique no line-shape model is required. In comparison to other phase-sensitive methods (PnP)~\cite{Cornell1989}, the phase evolution proceeds at notably small kinetic energies, leading to smaller systematic shifts.

\subsection{Determination of the Larmor frequency}
\begin{figure}[htb]
  \centering
  \includegraphics[width=0.8\textwidth]{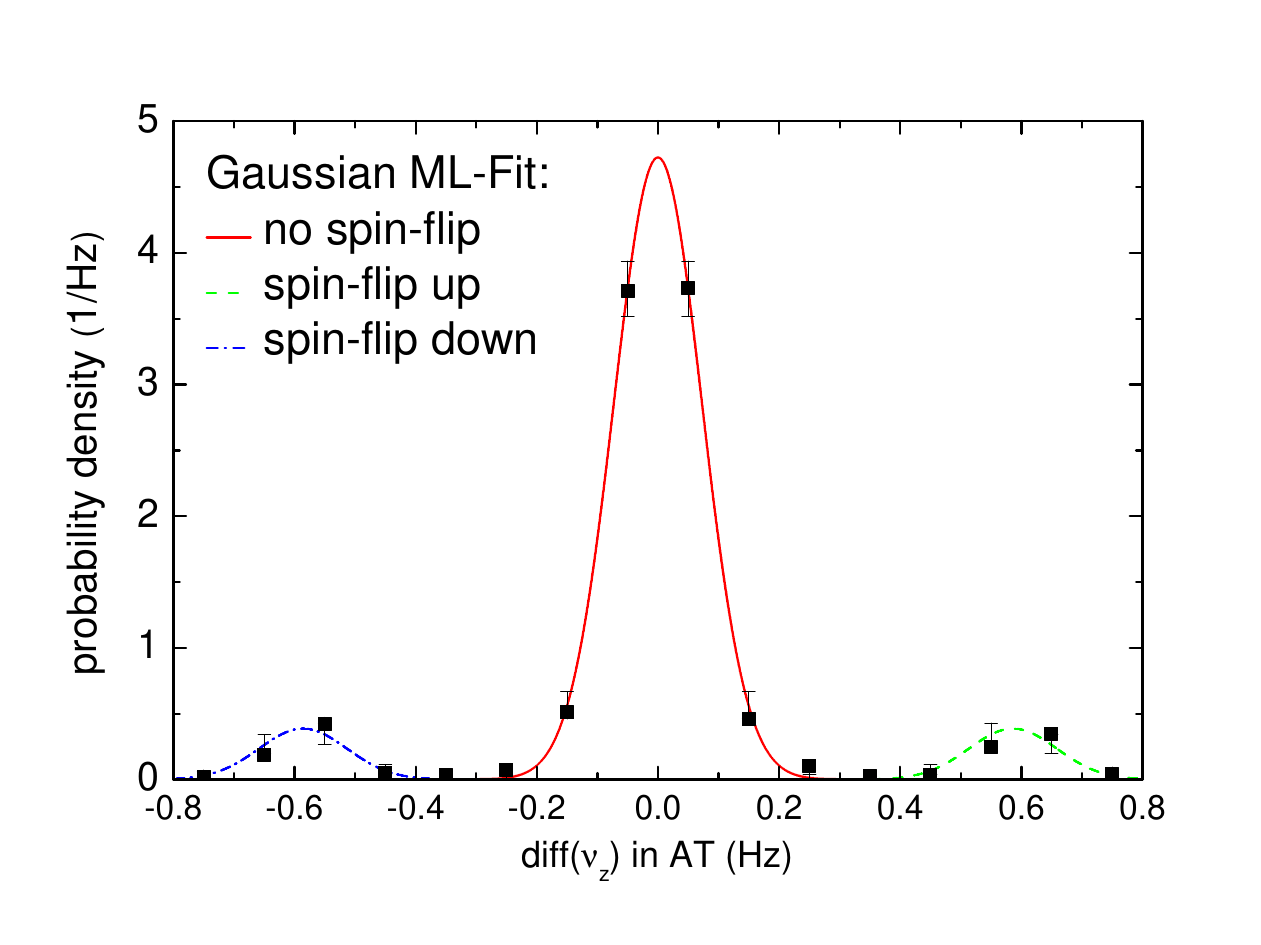}
  \caption{Probability density of the measured axial frequency differences in the analysis trap (AT), detecting the 
  axial peak signal. $\tN{diff}(\nu_z)$ is the axial frequency difference of subsequent measurements in the AT with $30\di{s}$ spin-flip drives in between. From a maximum-likelihood (ML) fit, which combines three Gaussian distributions (red: no spin-flip (sf), green: spin-flip up, blue: spin-flip down), the following parameters are extracted: spin-flip rate of 14.1$\%$, frequency jitter of $\delta \nu_z=73\di{mHz}$ and an axial frequency jump due to a spin-flip of $\Delta \nu_z^\tn{sf}=586\di{mHz}.$ The binned data (black markers) are just shown for visualization. The corresponding binomial error bars are determined from the number of measurement cycles per histogram-bin and the probability density at the bin center given by the ML fit.}
  \label{fig:EmassATvzJitter}
\end{figure}

Due to relative magnetic field fluctuations in the order of $5\cdot10^{-10},$ 
which are larger than the relative intrinsic Fourier width of about $1\cdot10^{-12}$ ($=\nu_\tn{Rabi}/\nu_L\approx0.1\di{Hz}/105\di{GHz}$), only an incoherent, Rabi-like excitation at the Larmor precession frequency is possible. For its determination we scan the frequency of a microwave field around the expected spin-flip frequency of $\nu_L\approx 105.43\di{GHz}.$ The field is generated by a microwave synthesizer (Anritsu MG3692B), locked to an atomic frequency standard, and a subsequent active frequency multiplier chain (OML S10MS). The detection of a spin-flip requires the knowledge of the spin-state before and after each attempt. This is performed by means of the \textit{continuous Stern-Gerlach effect}, which has been developed in 1986 by Hans Dehmelt~\cite{Dehmelt1986_ContinuousSternGerlach}. Here, a magnetic bottle-like field, $B_\tn{bottle}=B_2 z^2,$  in our case: $B_2=10.5(4)\cdot10^3 \di{T}/\tN{m}^2,$ produced by a ferromagnetic ring in the center of the trap, is added to the homogeneous Penning trap field, $B_0=3.7\di{T}$. This leads to a small spin-dependent harmonic axial force in addition to the electric trapping force. 
As a consequence the axial frequency depends on the spin-state due to the additional magnetic potential:
\begin{equation}
\label{eqn:MagnBottle}
\Phi_\tn{z, magn. bottle}=\mu_z B_2 z^2,
\end{equation}
where $\mu_z=-g \mu_B s_z /\hbar$ and $s_z=\pm \frac{\hbar}{2}.$
In the case of hydrogen-like carbon and at an absolute axial frequency of $\nu_z \approx 412\di{kHz}$ the axial frequency difference is: 
\begin{equation}
\label{eqn:dvzMagnBottle}
\Delta \nu_z^\tn{sf}=\frac{g \mu_B B_2}{m_\tn{ion} \omega_z}= 580(20)\di{mHz}.
\end{equation}
For the previously used \SiThirteen~ion~\cite{Sturm2011_hydrogenlikeSi} this shift was $\Delta \nu_z^\tn{sf} = 240 \di{mHz}.$\\
For the determination of the spin-state, we alternately measure $\nu_z$ and apply a $30\di{s}$ spin-flip drive, until we determine an axial frequency jump, which corresponds to a spin-flip. At this point, we measure the axial frequency by a $10 \di{ms}$ axial dipole excitation and a subsequent detection of the axial frequency peak signal. In addition, the resolution of the peak signal is improved by advanced FFT techniques, e.g. zero-padding \cite{ZeroPadding}. 
In figure~\ref{fig:EmassATvzJitter} the distribution of the axial frequency differences of the subsequent measurements is shown. From a Gaussian maximum likelihood (ML) fit, we extract an axial frequency jitter of $\delta \nu_z=73\di{mHz}$, a spin-flip probability of  $14.1\%$ and an axial frequency jump of $\Delta \nu_z^\tn{sf} = 586\di{mHz}$ in agreement with the expectation. The spin-state can be resolved to $99.995\%$ (4 $\sigma$ confidence) which corresponds to: $\Delta \nu_z^\tn{sf}/2 = 4.04  \cdot \delta \nu_z.$ 

\subsection{The triple Penning trap setup}
\begin{figure}[htb]
\centering
\includegraphics[width=0.7\textwidth]{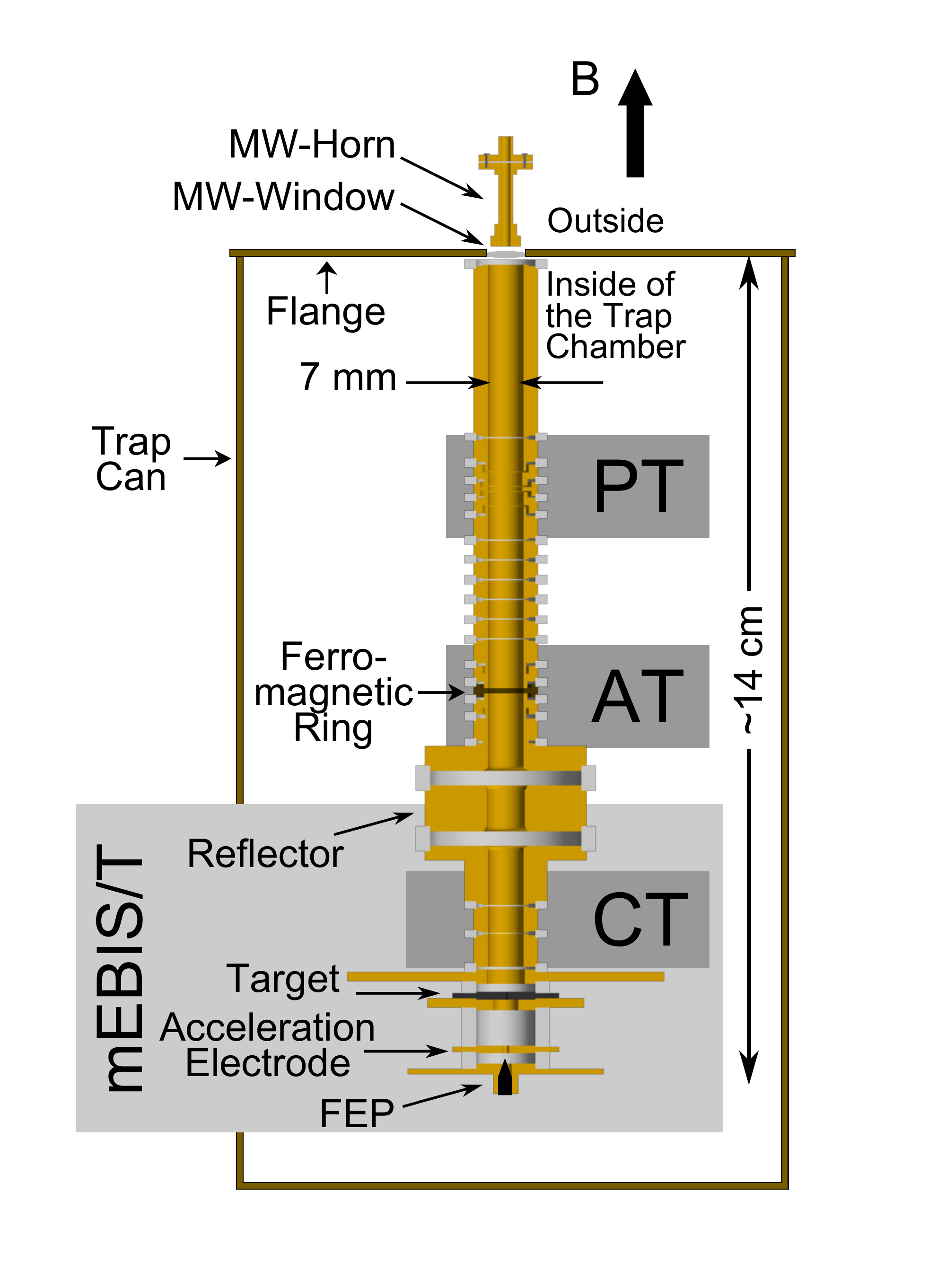}
\caption{\label{fig:ExpSetupII}Trap tower, including precision trap (PT), analysis trap (AT) and creation trap (CT), placed in a  vacuum vessel.}
\end{figure}

The requirements of a strongly inhomogeneous magnetic field for the spin-state detection on the one hand and an as perfect as possible field homogeneity for precise eigenfrequency and Larmor frequency measurements on the other hand are conflicting. Therefore we use two spatially separated Penning traps with a distance between the trap centers of $41\di{mm},$ see figure~\ref{fig:ExpSetupII}. 
In the so-called \textit{analysis trap} (AT) a ferromagnetic ring electrode with a saturation flux density of $B_\tn{s}=0.645\di{T}$ (nickel) induces the magnetic bottle for the spin-state detection as mentioned in the previous paragraph. The \textit{precision trap} (PT) has a homogeneous magnetic field for the detection of the eigenfrequencies and the probing of the Larmor precession frequency. The leading magnetic field imperfections in the PT are: $B_1=-13.41(23)\cdot 10^{-3}\di{T/m}$ and $B_2=1.01(0.20) \di{T/m}^2$ \cite{Sturm_PhD}. Assuming the magnetized ring electrode in the AT as the only source of magnetic field inhomogeneities, these higher-order magnetic field coefficients are in good agreement with finite element simulations using \Comsol ~\cite{Sturm_PhD,comsol}.\\
The trap tower is enclosed in a cold-welded cryogenic chamber, also made from OFHC copper. An indium-sealed flange contains the voltage, signal and excitation lines and a quartz glass window, which enables the coupling of the microwaves for the probing of the Larmor precession frequency. Based on theoretically predicted electron transfer cross sections~\cite{Hermanspahn_PhD} and ion's storage time of several months, we estimate the pressure to $\approx 10^{-17}\di{mbar},$ which corresponds to less than 20 hydrogen molecules or helium atoms in the trap volume. Other atomic or molecular species are frozen out at these cryogenic temperatures.\\
A miniature electron beam ion source and trap (mEBIS/T) is used for the ion production \cite{Alonso_2006}. It contains a third 3-electrode Penning trap, the \textit{creation trap} (CT), see figure~\ref{fig:ExpSetupII}.

\subsection{Creation of a single highly charged ion}
For the ion creation process an electron beam of a few $100\di{nA}$ is emitted by a field emission point (FEP)~\cite{Hermanspahn_PhD}. The beam impinges on a target, desorbes and subsequently ionizes different atomic  species. The ions are trapped in the CT. Consecutive ionization by the electron beam produces ions in higher charge states. The energy of the electron beam is well above the ionization threshold of hydrogen-like carbon, $E_\tn{ion}=392 \di{eV}.$ 
The time period of the complete creation process for a cloud of highly charged ions is in the order of $10\di{s}.$ Afterwards the complete cloud is transported from the CT to the PT.\\
The most efficient way to remove unwanted ion species is the highly selective $B_2$\textit{-cleaning} technique: (1) Via a short radial dipole excitation at the expected modified cyclotron frequency of \CFive~we exclusively excite the modified cyclotron motion of all \CFive~ions in the PT. (2) Subsequently, we transport this modified ion cloud into the AT. The axial mode of the \CFive~ions is cooled to $\approx 5\di{K}$ via the axial resonator. (3) Lowering the electric trapping potential below the thermal axial energy, the unwanted ion species are evaporated, while the \CFive~ions remain confined, due to the force acting on their large orbital magnetic moments: $\mu_z^\tn{cycl}=0.5 q  r_+^2\omega_+$ by the magnetic bottle $B=B_2 z^2.$ In order to keep the \CFive~ions in the trap, the axial magnetic energy $E_z^\tn{cycl}=\mu_z^\tn{cycl} B_2 z^2$ must be larger than the axial thermal energy $E_z=k_B T_z$. (4) The remaining \CFive~ions are transported back from the AT into the PT, where (5) their modified cyclotron mode is cooled.\\
Multiple \CFive ~ions with slightly different motional energies can be distinguished by their different modified cyclotron frequencies detected in the PT as different peak signals in the Fourier spectrum of a cyclotron resonator with low quality factor $Q.$ These ions can be selectively removed by a cautious lowering of the trapping potential until a single ion remains. The characteristic parameters of the cyclotron resonator (resonance frequency, quality factor and cooling time constant for a resonant \CFive ~ion) are listed in table~\ref{tab:CyclotronReso}.
\begin{table}
\centering
\caption{\label{tab:CyclotronReso}Characteristic parameters of the cyclotron resonator in the PT.}
\footnotesize	
\begin{tabular}{l c }
\br
cyclotron resonator & value\\
parameters: & \\
\mr
$Q$ & $120(10)$\\
$\tau(\nu_+)$ & $415(100) \di{s}$\\
$\nu_\tn{res}$ & $24\;455(15) \di{kHz}$\\
$\nu_+$ & $24\;081\;134 \di{Hz}$\\
\br
\end{tabular} 
\normalsize
\end{table}

\subsection{The measurement process}
The measurement cycle is started in the AT. Here, measurements of the ion's axial frequency ($40\di{s}$) and in between attempts to induce a flip of the electron spin ($30\di{s}$) are repeated until a jump in the axial frequency, arising from the continuous Stern-Gerlach effect, is observed. Each time $\nu_z$ is measured by averaging over six zero-padded axial peak signals. This procedure leads to the knowledge of the spin state. Afterwards the ion is transported from the AT to the PT. Here, the electrostatic trapping potential is much slower changed compared to the ion's oscillation periods, resulting in an adiabatic transportation of the ion.\\
In the PT, we measure at first the modified cyclotron frequency $\nu_+^\tn{DD}$ with a double-dip (DD) signal (averaging time: $200\di{s}$), followed by a first axial frequency measurement $\nu_z^I$ via a dip-signal (averaging time: $200\di{s}$). Then, the PnA-cycle is realized 10 times with the following phase evolution times: $6 \times 10\di{ms}, 1\di{s}, 2\di{s}, 2 \times 5\di{s}.$ The $10\di{ms}$ measurements determine the starting phase. They can be repeated to reduce the uncertainty of the starting phase, since magnetic field fluctuations can be neglected at these time scales. The double-dip measurement, as well as the $1\di{s}$ and $2\di{s}$ PnA-cycles, are essential for a proper phase unwrapping. The first $5\di{s}$ measurement determines the current modified cyclotron frequency and thus the magnetic field. From this information the scanned frequency $\nu_\tn{MW}$ of the microwave field is chosen, which is injected during the second $5\di{s}$ measurement. 
Since the probing of the Zeeman transition at  $\nu_\tn{MW}$ and the measurement of the modified cyclotron frequency, $\nu_+^\tn{PnA}$, happen at the same time during the last PnA-cycle, magnetic field fluctuations cancel to a large extent in the ratio of the frequencies. 
\begin{figure}[htb]
  \centering
  \includegraphics[width=0.6\textwidth]{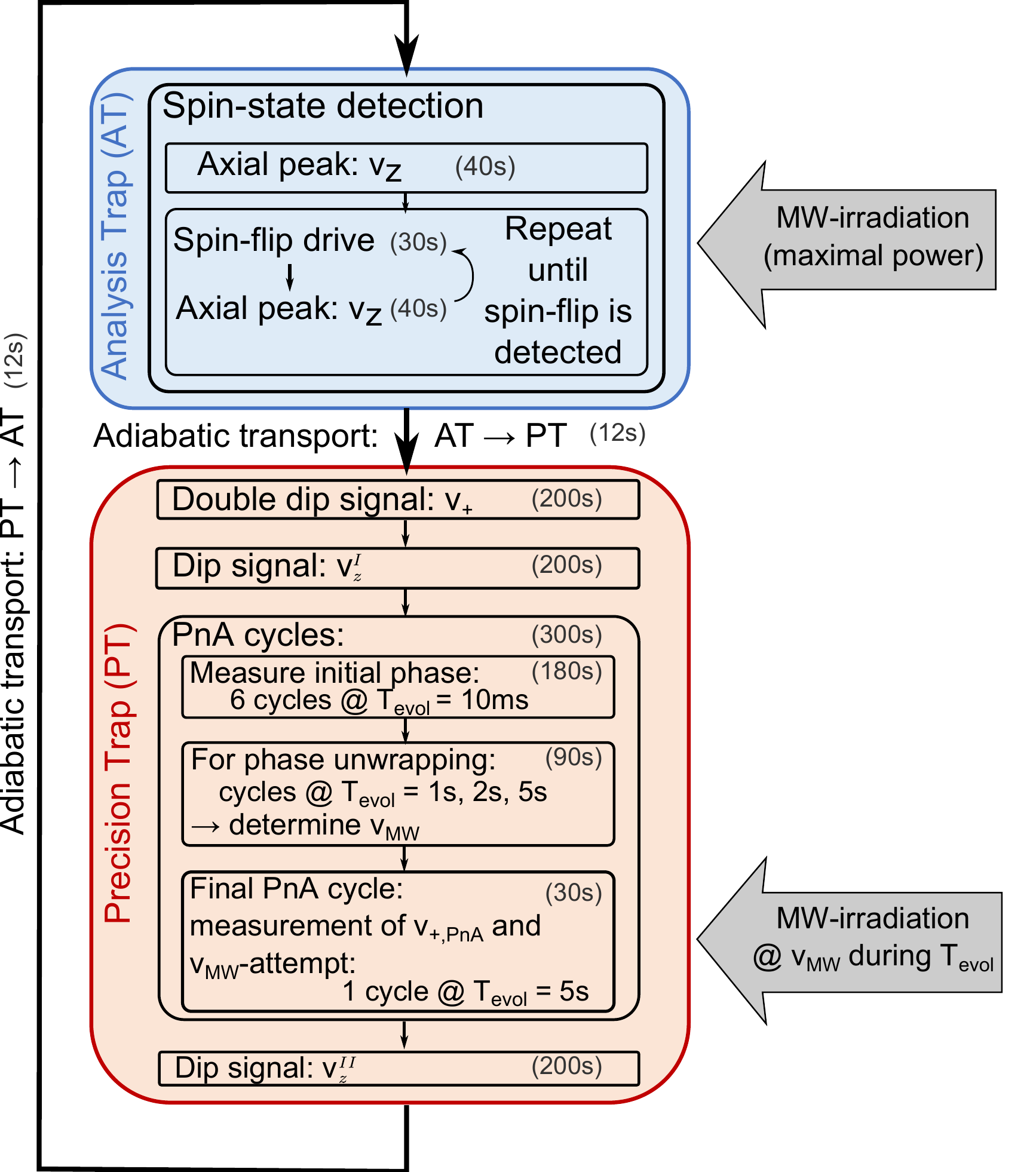}
  \caption{  \label{fig:MeasScheme}Flow chart of a measurement cycle. For details see text.}
\end{figure}To consider potential changes of the trapping voltage during the PnA-cycles a second dip-signal $ \nu_z^{II}$ is measured for an interpolation of the axial frequency, $\nu_z=(\nu_z^I+\nu_z^{II})/2$. From $\nu_+^\tn{PnA},$ $\nu_z$ and $\nu_-$ from previous measurements\footnote{The contribution of $\nu_-$ to $\nu_c$ is very small and only an occasional measurement is required.}  we calculate the free cyclotron frequency: $\nu_c=\sqrt{\nu_{\tn{PnA},+}^2+\nu_z^2+\nu_-^2}.$ In combination with the probed Larmor frequency $\nu_\tn{MW}$ we determine the frequency ratio: $\Gamma^*=\nu_\tn{MW}/\nu_c$.\\
Finally the ion is transported back to the AT and the spin state is determined in the same way as at the beginning of the cycle in order to know whether the attempt to flip the spin in the PT was successful or not. The whole cycle, which is also illustrated in figure~\ref{fig:MeasScheme}, takes typically $25\di{min}.$\\
The recording of a complete $\Gamma$-resonance 
\begin{figure}[htb]
  \centering
  %
    \includegraphics[width=0.8\textwidth]{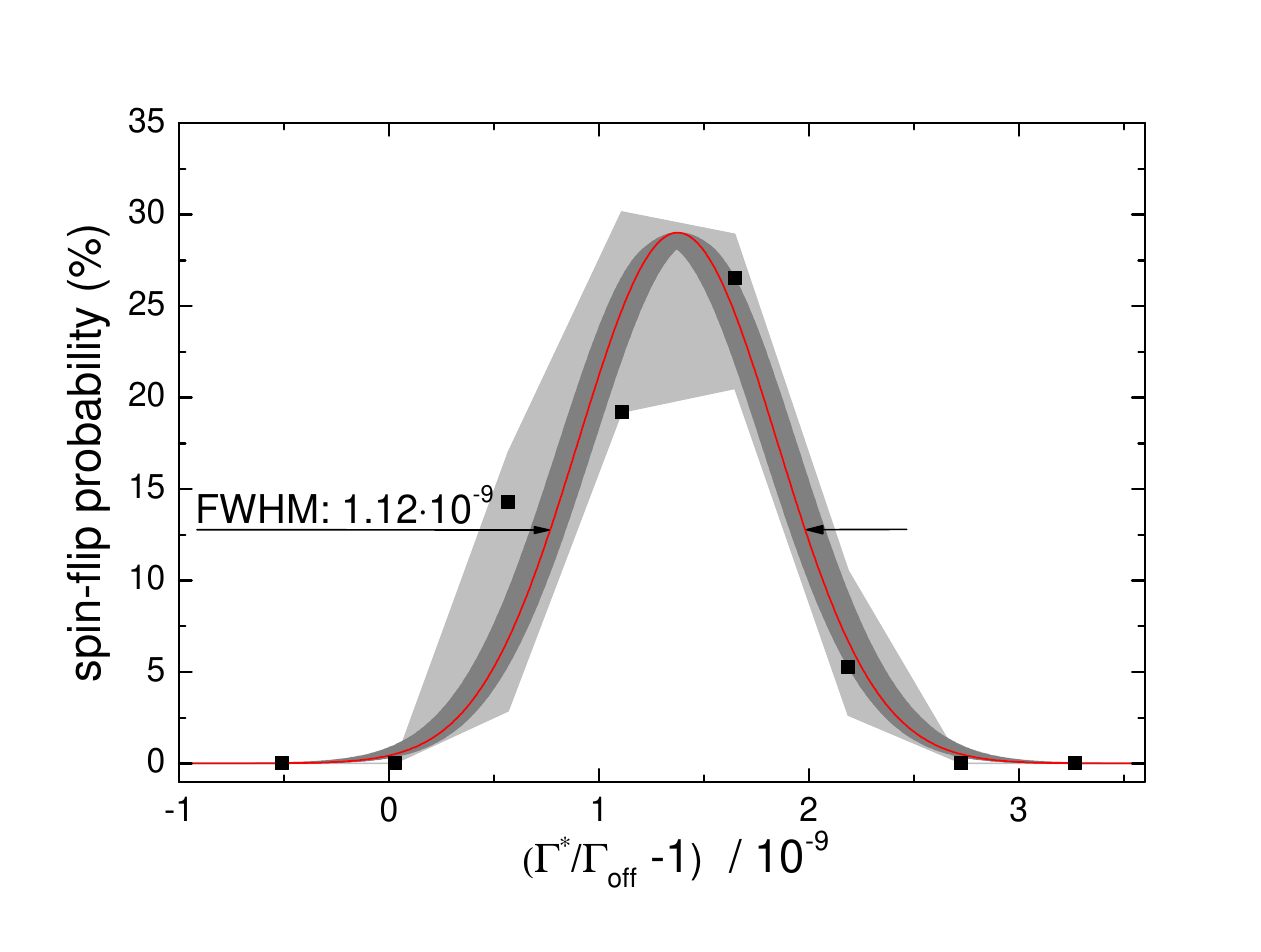}
\caption{  \label{fig:GammaReso}A $\Gamma$-resonance represents the spin-flip rate in the PT in dependence of the measured frequency ratio $\Gamma^*.$ Here, the frequency ratio is scaled by a constant: $\Gamma_\tn{off}=4\;376.210\;497\;791$. To guide the eyes, randomly histogrammed data are indicated by the black points. The final $\Gamma$ value is extracted from the maximum-likelihood (ML) fit-routine shown in red line. The corresponding line-shape model is discussed in section~\ref{sec:lineShape}. The dark gray area covers all possible ML line-shapes within the 1$\sigma$ uncertainty of the final $\Gamma$ value. The bright gray area illustrates the binomial uncertainty of the histogrammed data based on the spin-flip probability of the ML-fit. The resonance includes 43 measurement cycles which feature a spin-flip in the PT and 236 measurement cycles featuring no spin-flip in the PT. 
}
\end{figure}which exhibits the spin-flip rate in dependence of the measured frequency ratio $\Gamma^*$ requires several 100 frequency ratio measurements and the respective binary spin-flip information (spin has / has not flipped). Figure~\ref{fig:GammaReso} shows the results of 279 attempts to induce a spin-flip at the measured frequency ratios $\Gamma^*$. Here, the modified cyclotron energy amounts to $E_+=4.2(5)\di{eV}$ ($r_+^\tn{evol}= 54(3)~\mu\tN{m}$) during the PnA phase evolution time of $T_\tn{evol}=5\di{s}.$ The resonance is fitted by a line-shape discussed below. Since the phase jitter of the modified cyclotron mode represents the main contribution to this line-shape, we discuss it separately in the following chapter.

\section{Sources of the modified cyclotron phase jitter}\label{sec:jittersources}
\begin{figure}
\subfloat[]{\includegraphics[width=0.49\textwidth]{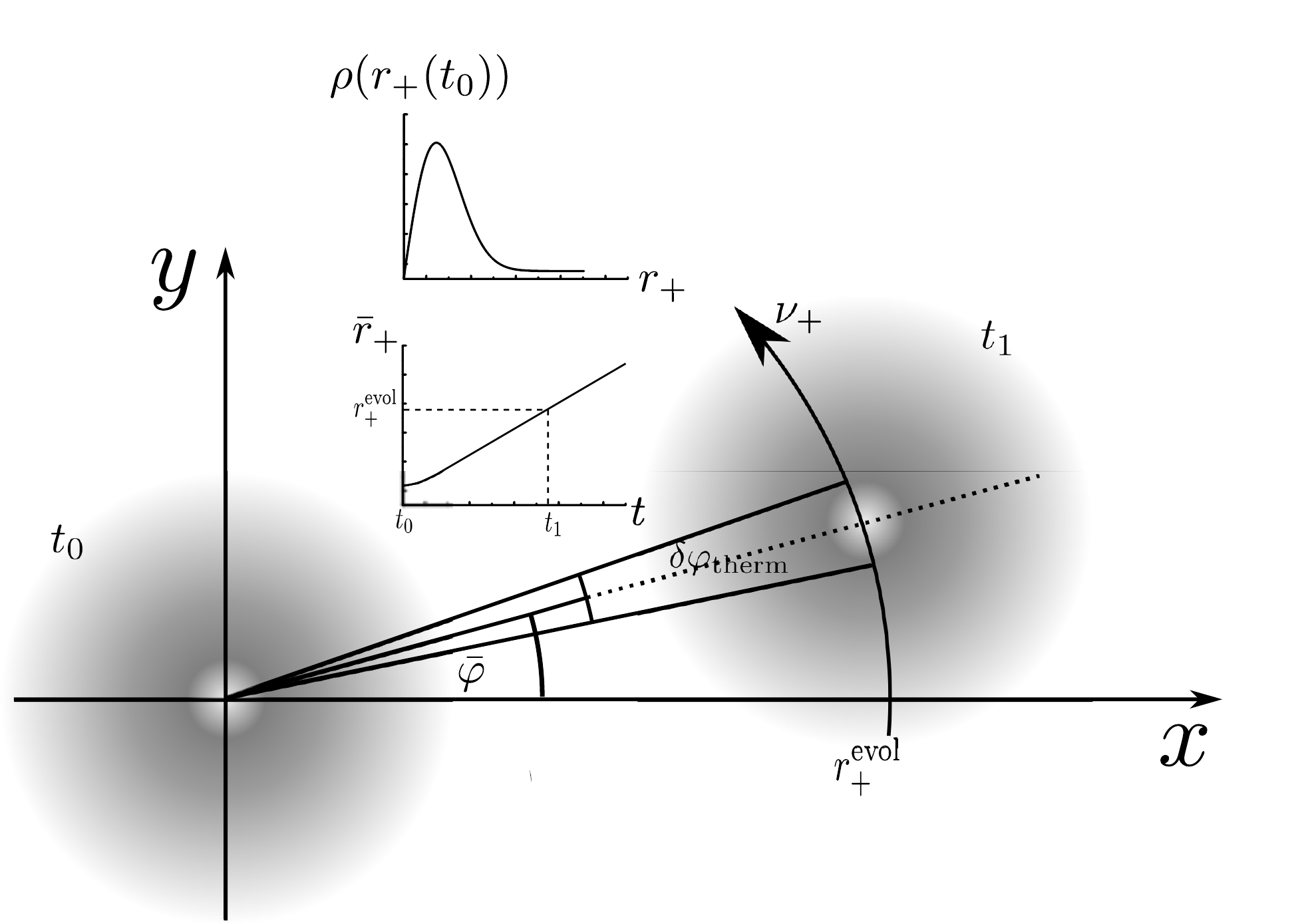}}\hfill
\subfloat[]{\includegraphics[width=0.49\textwidth]{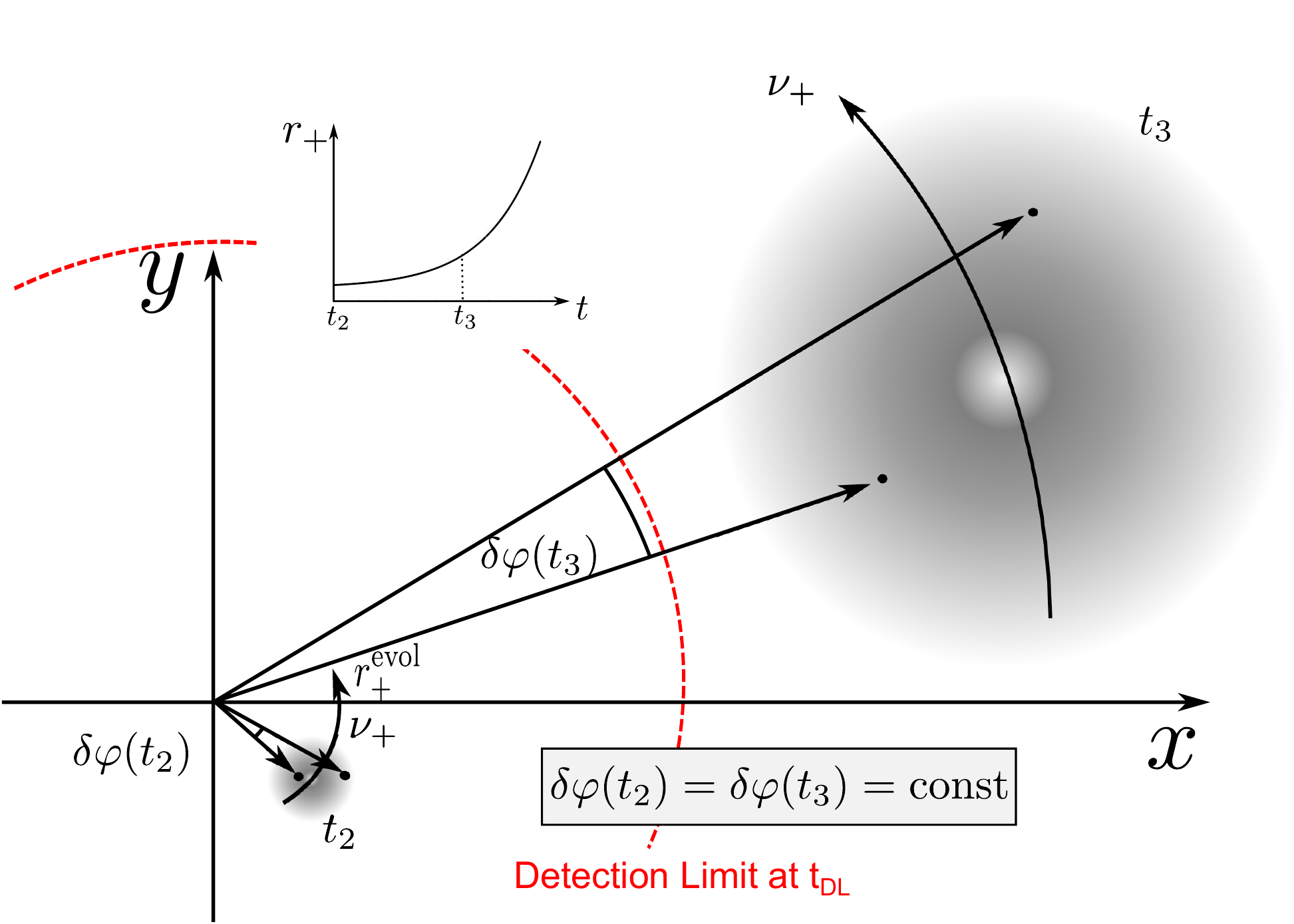}}
 \caption{\label{fig:PnAPulses}Illustrations of the modified cyclotron radius distributions, indicated as dark clouds, in the radial (x/y) plane of the trap before and after the first (a) and the second (b) PnA pulse. (a) During the first PnA pulse, the spread of the modified cyclotron radius is kept constant. The upper inset illustrates the Boltzmann distribution of the modified cyclotron radius. The lower inset shows the linear increase of the modified cyclotron radius during this dipole excitation. (b) During the second PnA pulse the modified cyclotron radius increases exponentially, which is presented in the inset. Here, the radial spread of the modified cyclotron radius distribution scales with $r_+$, so that no additional phase jitter arises: $\delta \varphi(t_2)=\delta \varphi(t_3).$ For further details see text.}
\end{figure}

\begin{figure}[htb]
  \centering
  \includegraphics[width=0.7\textwidth]{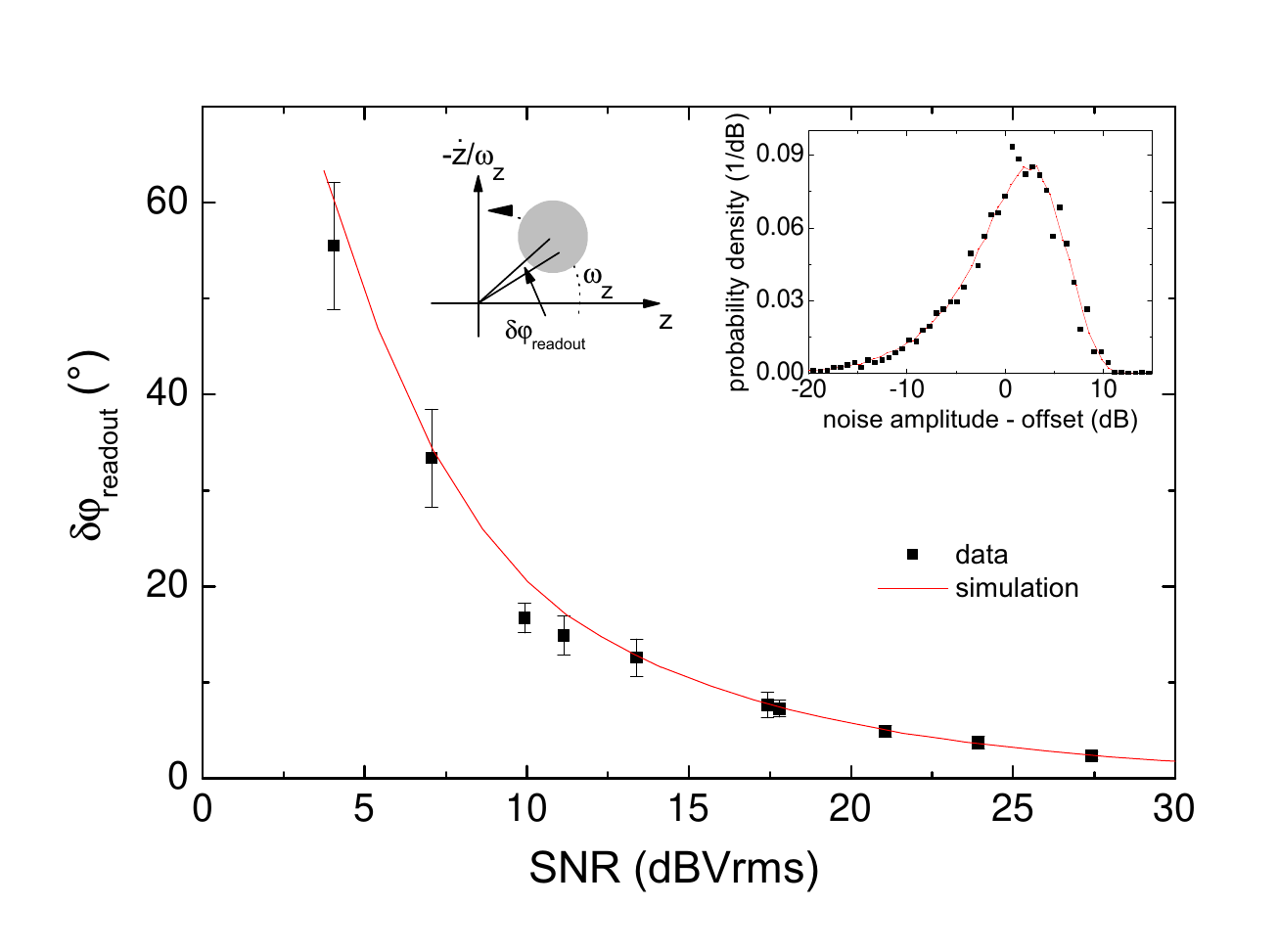}
  \caption{Measurements of the intrinsic technical phase detection uncertainty by an artificially generated, exponentially decreasing sine signal ($\tau= 260\di{ms}  \approx 0.5\cdot \tN{signal length}$) for different signal-to-noise (SNR) ratios. 
The numerically predicted detection uncertainty (red line) is in reasonable agreement with the data. This prediction relies on the distribution of the noise amplitudes, see gray cycle in the left inset, which is also in good agreement with the expectation, see right inset.}
  \label{fig:ReadOutJitter}
\end{figure}

\begin{figure}
\subfloat[]{\includegraphics[width=0.49\textwidth]{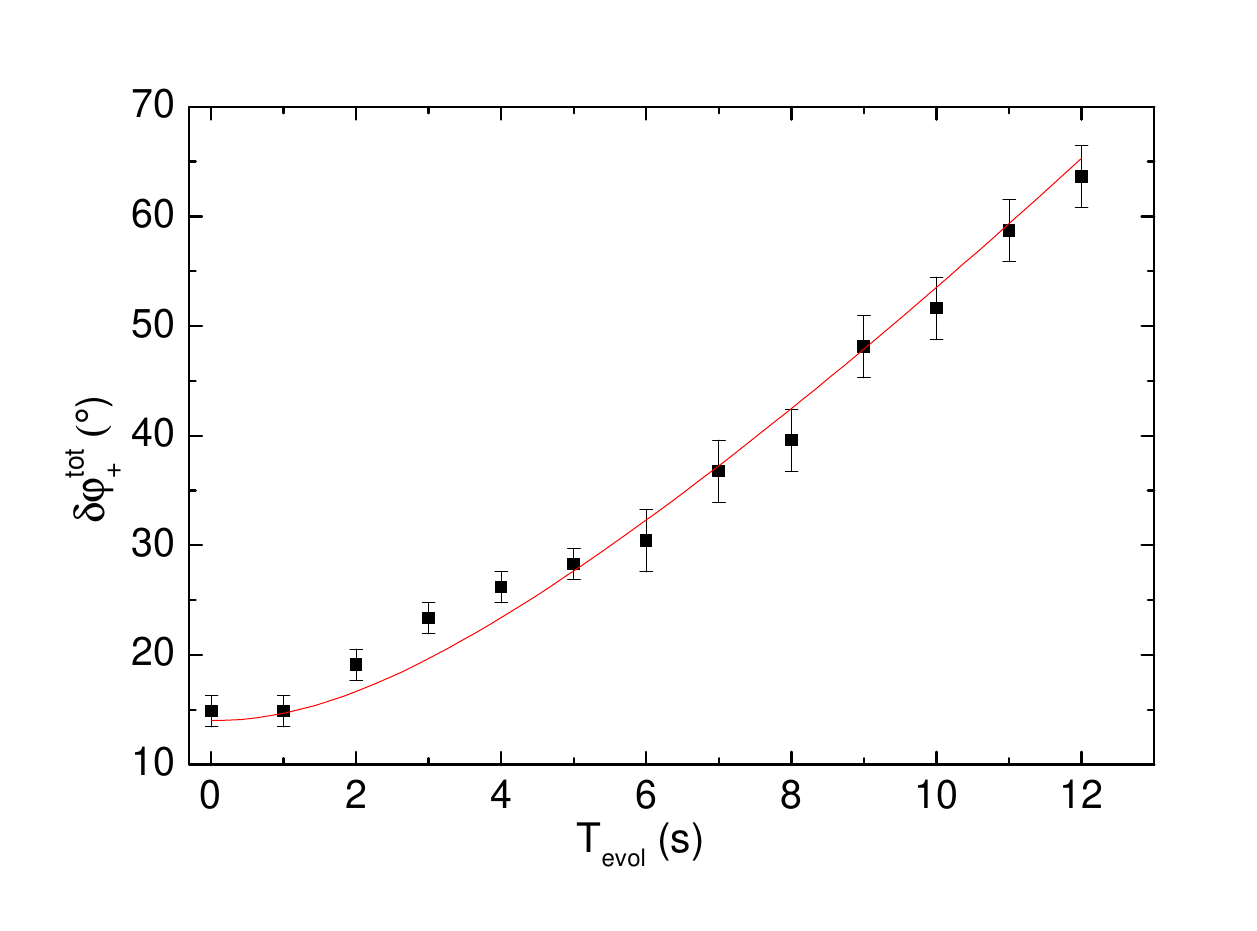}}\hfill
\subfloat[]{\includegraphics[width=0.49\textwidth]{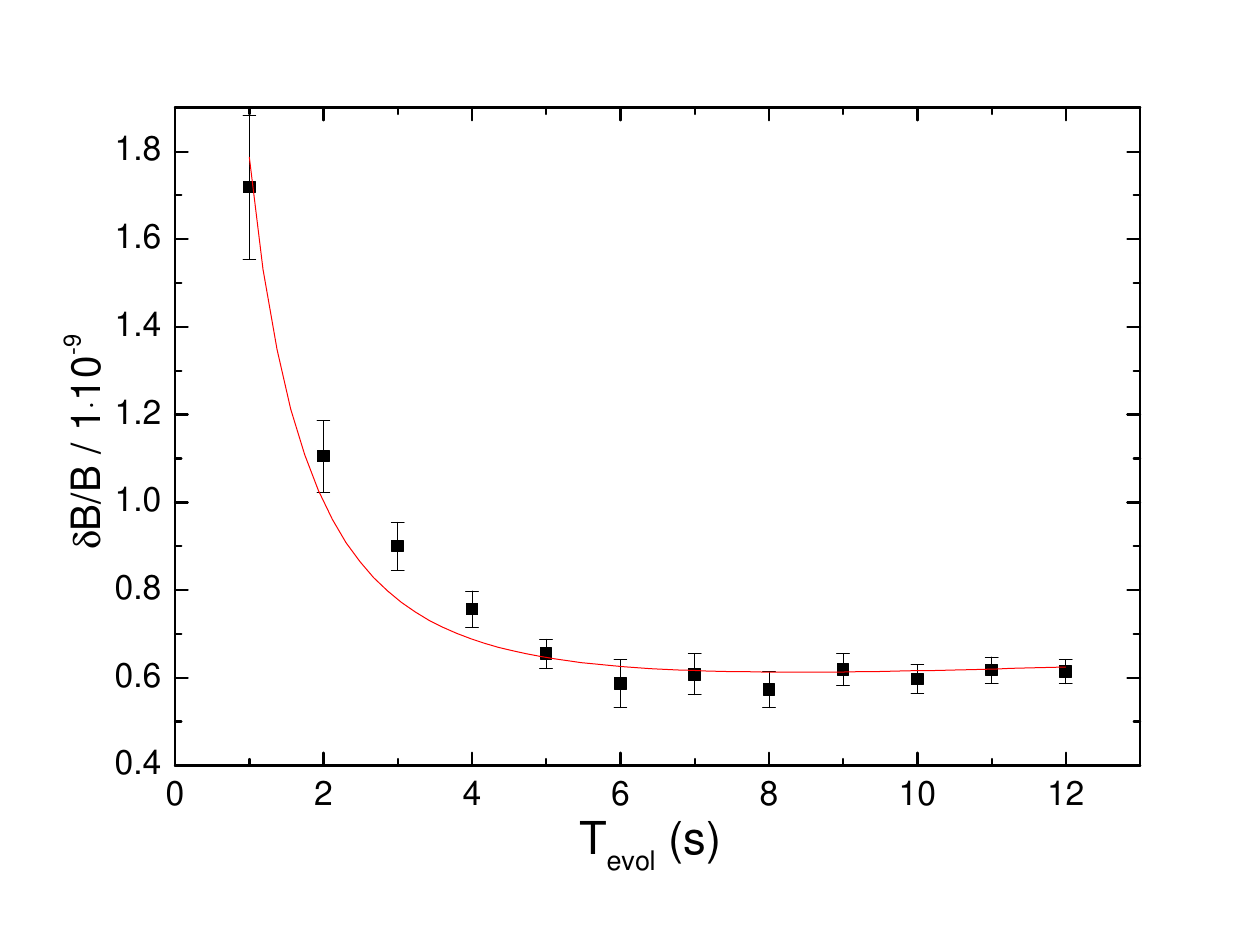}}
  \caption{ \label{fig:MagneticFieldJitter}Subsequent PnA measurements of the magnetic field jitter in the PT at $r_+^\tn{evol}=30(2)\;\mu\tN{m}$. Plot (a) presents the measured phase jitter versus the phase evolution time and (b) the corresponding relative jitter of the modified cyclotron frequency. The fit functions (red lines) rely on a random walk model of the magnetic field.}
\end{figure}

The line-shape model of the $\Gamma$-resonance requires a detailed understanding of the composition of the modified cyclotron phase jitter\footnote{As a phase jitter we define the standard deviation of subsequently measured phase differences  divided by the square root of two: $\delta \varphi=\tN{std}(\tN{diff}(\varphi))/\sqrt{2},$ which is often also termed as Allan deviation.}. Three different sources of phase jitter have to be considered: An intrinsic thermal phase jitter, an intrinsic ''technical'' readout-related phase uncertainty and the magnetic field jitter. In the following these three types of jitter are studied separately.\\
{\bf (1) The intrinsic thermal phase jitter} is characterized by the primarily thermalized energy distribution of the modified cyclotron mode and the first PnA pulse. During this radial dipole excitation at $\nu_\tn{rf}\approx\nu_+$ the spread of the modified cyclotron radius is kept constant. The modified cyclotron radius itself increases linearly up to $r_+^\tn{evol}=13\; \mu\tN{m}$ or larger. The modified cyclotron radius distributions are illustrated in figure~\ref{fig:PnAPulses}(a) as dark clouds in the radial (x/y) plane before ($t_0$) and after ($t_1$) the first PnA pulse. In principle, this jitter can be further reduced by a better, improved cooling of the initial modified cyclotron mode or by an increase of the excitation strength 
of the first pulse.
The latter way, however, is not favorable, since higher modified cyclotron energies during the phase evolution time lead to larger systematic shifts with increased uncertainties. After the phase evolution time $T_\tn{evol}=t_2-t_1$ the second PnA pulse, a parametric excitation at $\nu_\tn{rf}\approx \nu_++\nu_z,$ is performed. In figure~\ref{fig:PnAPulses}(b)  the modified cyclotron radius distributions are shown before ($t_2$) and after ($t_3$) this pulse. Here, $r_+$ increases exponentially and after a time $t_\tn{DL}$ it exceeds the detection limit. If $r_+^\tn{evol}>z \sqrt{\nu_z/\nu_+}=3.5\;\mu\tN{m}$ at the beginning of the second PnA pulse, the phase-spread of the thermal distribution scales with $r_+$, so that no additional phase jitter arises: $\delta \varphi(t_2)=\delta \varphi(t_3).$\\
{\bf (2) The intrinsic technical phase detection uncertainty} arises during the readout process of the peak signal. Since it is completely independent of the motion of the ion, we studied this effect by inducing an artificial peak signal a few kHz next to the axial resonator of the PT ($\approx 630\di{kHz}$) via the quadrupole excitation. 
The peak signal has been generated by an exponentially decreasing sine signal, simulating a thermalizing excited ion. The readout signal has the usual length of $512 \di{ms}$. The exponential modulation has a decay constant of $260 \di{ms},$ similar to the cooling time constant of the \CFive ~ion. The technical jitter scales with the signal-to-noise ratio (SNR) of the peak signal, see figure~\ref{fig:ReadOutJitter}. The numerically predicted detection uncertainty (red line) is in reasonable agreement with the data.
An unfavorable axial peak position with respect to the FFT-binning, the so-called frequency bleeding, might reduce the apparent SNR. In case the peak-signal \textit{bleeds} likewise into two FFT-bins, the SNR is reduced by maximally $3 \di{dB}.$ 
The phase detection uncertainty can be diminished by a larger SNR, which in the future might be achieved by a resonator with a higher quality factor or by a larger excitation during the second PnA pulse. The excitation strength is limited by higher-order electric field imperfections~\cite{Ketter20141} in combination with the initial thermal distribution of the amplitudes, which shift the axial frequency out of the readout bin.\\
The combination of the thermal phase jitter and the phase detection uncertainty has been studied in PnA-cycles with short phase evolution times of $10 \di{ms}$.
\begin{figure}
\subfloat[]{\includegraphics[width=0.49\textwidth]{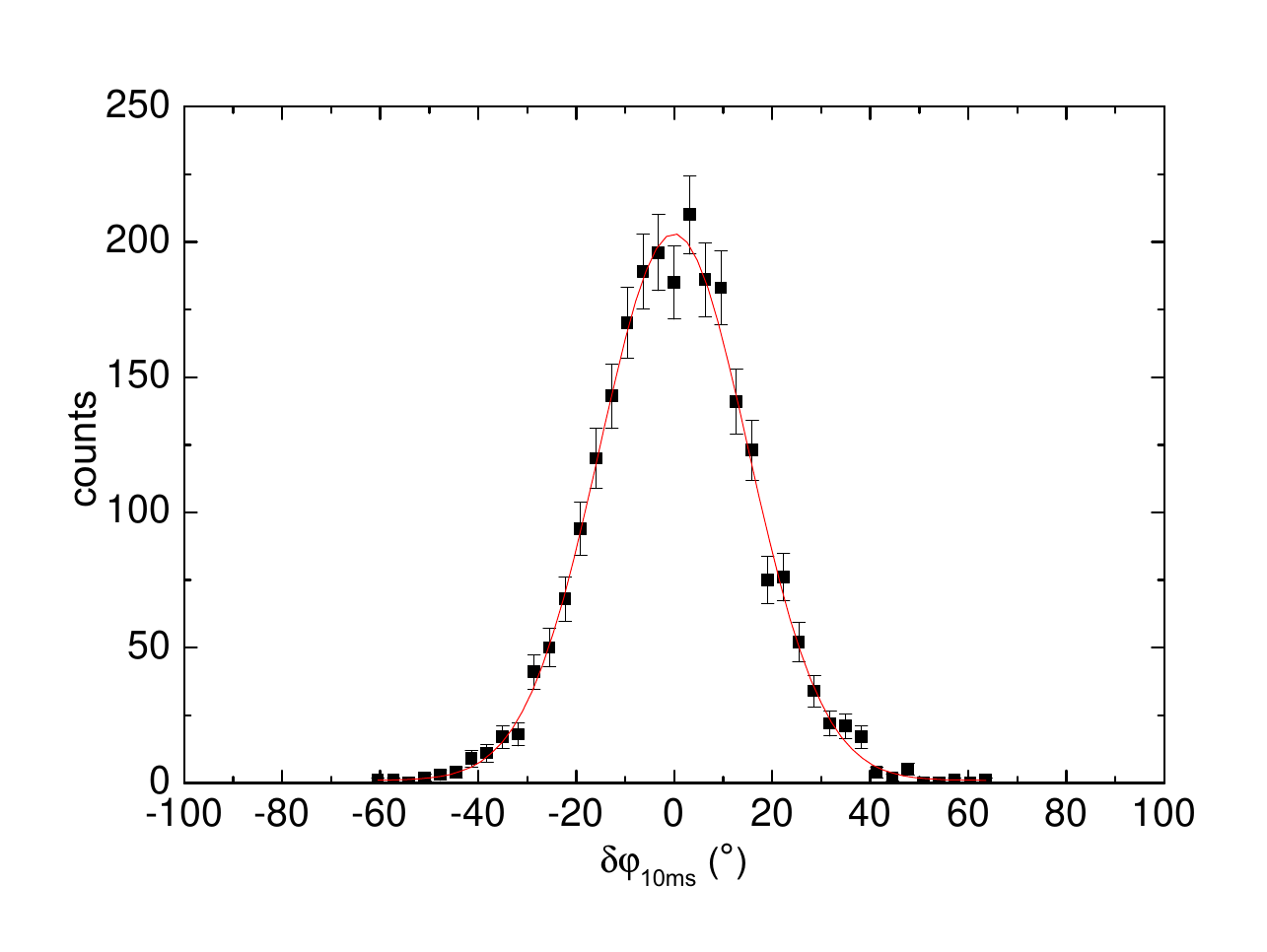}}\hfill
\subfloat[]{\includegraphics[width=0.49\textwidth]{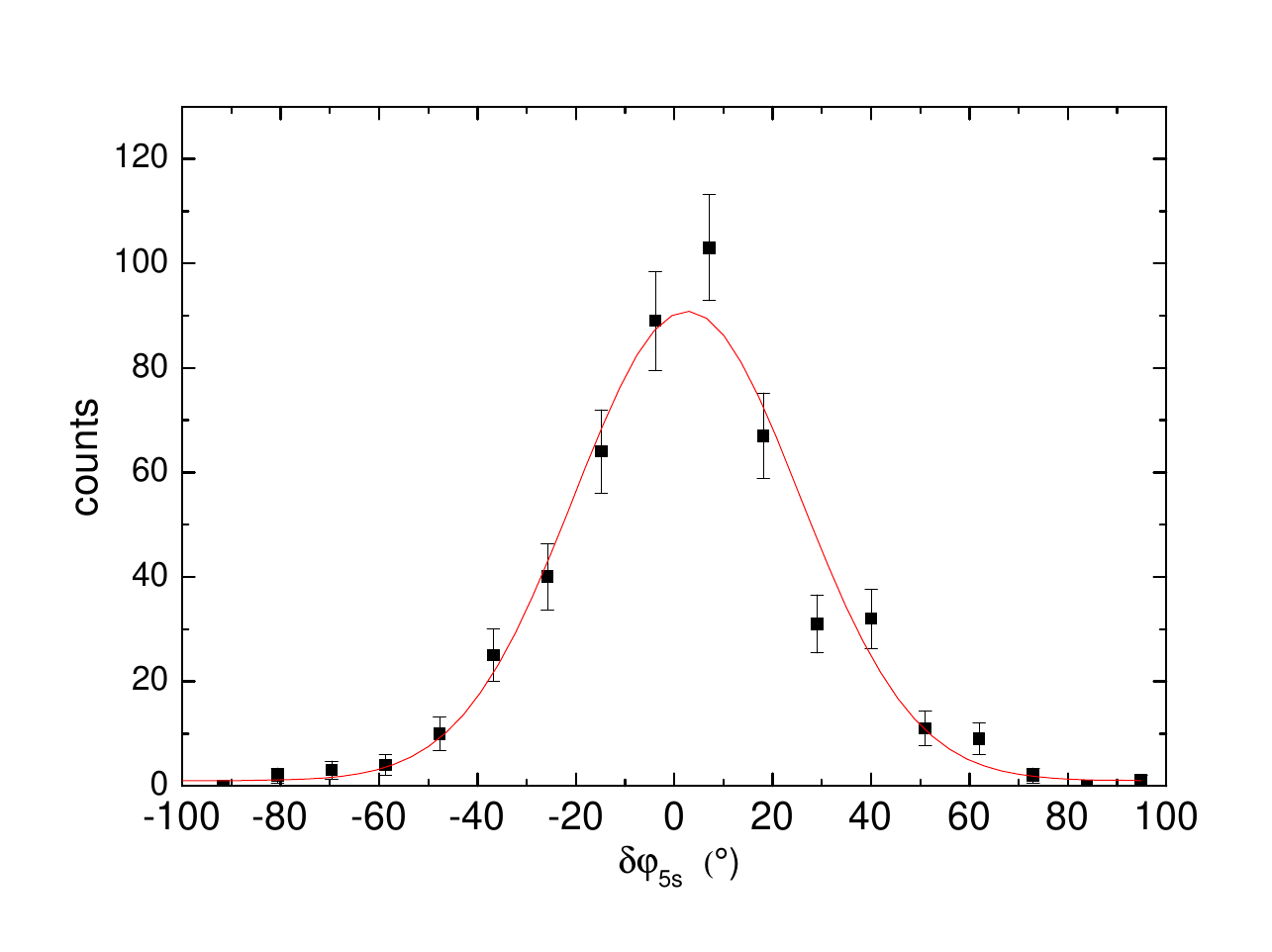}}
  \caption{\label{fig:DistributionsPnA}Histogrammed phase differences of subsequent PnA-cycles ($\delta\varphi=\tN{diff}(\varphi_z)/\sqrt{2}$) at small modified cyclotron energies ($r_+^\tn{evol}=30(2) \; \mu\tN{m}$) with phase evolution times of $T_\tn{evol}=10\di{ms}$ (a) and $T_\tn{evol}=5\di{s}$ (b). Both distributions have a Gaussian line-shape (red line).}
\end{figure}In figure~\ref{fig:DistributionsPnA}(a) the Gaussian distributed phase differences of subsequent PnA measurements are histogrammed for small modified cyclotron energies during the phase evolution time ($r_+^\tn{evol}=30(2) \; \mu\tN{m}$).
\begin{table}
\centering
\caption{\label{tab:10msPhaseJitter}Analysis of the $10 \di{ms}$ phase jitter for \CFive. Comparison between the modeled phase jitter (the squared sum of the thermal jitter and the readout jitter) and the measured phase jitter in dependence of the radius and the measured SNR.}
\footnotesize
\begin{tabular}{ l l | c c c | c}
\br
 $r_+^\tn{evol} (\mu\tN{m})$  & avg-SNR & readout & thermal  & modeled & measured\\
 & (dB) & jitter ($^\circ$) & jitter ($^\circ$) & jitter  ($^\circ$) & jitter ($^\circ$)\\
 \mr
13(1) & 13.4(1) & 13.8(1.7) & 9.0(3) & 16.5(1.5) & 23.3(5)\\
30(2) & 13.0(2) & 14.6(1.8) & 3.9(2) & 15.1(1.7) & 15.8(5)\\
36(2) & 15.3(2) & 10.3(1.3) & 3.4(2) & 10.9(1.2) & 14.2(1.0)\\
54(3) & 11.0(2) & 18.2(2.2) & 2.2(2) & 18.3(2.2) & 16.6(1.1)\\
90(4) & 15.8(3) & 12.4(1.5) & 1.3(1)  &  12.4(1.5) & 11.8(1.1)\\
\br
\end{tabular}\\
\end{table}
\normalsize 
In table~\ref{tab:10msPhaseJitter} the measured phase jitter (6th column) is listed in dependence of the modified cyclotron radius (1st column, determined by the first PnA pulse) and the measured averaged SNR (2nd column, determined by the second PnA pulse). The modeled phase jitter (5th column), which is the squared sum of the calculated phase detection uncertainty (3rd column) and the calculated thermal jitter (4th column), is in reasonable agreement with the data. 
Significant deviations only occur at small radii.\\
{\bf (3) The magnetic field related phase jitter} is caused by magnetic field fluctuations during the phase evolution time. The time dependent behavior of these fluctuations can be modeled by a random walk of the magnetic field; the standard deviation is given by: 
\begin{equation}
\label{eqn:RandomWalk}
\delta B\propto \delta \nu_+ \propto\sqrt{T_\tn{cycle}},
\end{equation}
where $T_\tn{cycle} \approx T_\tn{evol}+T_\tn{cooling}$ is the time period of a complete PnA-cycle, mainly composed of the phase evolution time $T_\tn{evol}=10\di{ms}$ to $5\di{s}$ and the cooling time $T_\tn{cooling}=25\di{s}$ of all eigenmotions. In combination with the model of the $10\di{ms}$ phase jitter (detection uncertainty $+$ thermal jitter), the total phase jitter has the following form:
\begin{eqnarray}
\label{eqn:MagneticFieldJitter}
\delta \varphi_+^\tn{tot}&=\tN{std}(\tN{diff}(\varphi_+^\tn{meas}))/\sqrt{2}\nonumber\\
&=\sqrt{2(\delta \varphi_+^\tn{10ms})^2+(A \sqrt{T_\tn{cycle}} \cdot 360^\circ \cdot T_\tn{evol})^2}/\sqrt{2}.
\end{eqnarray}
In figure~\ref{fig:MagneticFieldJitter}(a) the total phase jitter of the modified cyclotron mode is plotted versus the phase evolution time of the PnA method. From these data the amplitude, $A=0.00339(7)\di{s}^{-1.5},$ of the random walk is extracted. The fit-function is shown in red. In figure~\ref{fig:MagneticFieldJitter}(b) the corresponding relative magnetic field jitter: $\delta B/B=\delta \varphi_+^\tn{tot}/\nu_+/360^\circ/T_\tn{evol}$  is shown. The optimal phase evolution time for a minimal frequency jitter is $8.3\di{s}$. 
It should be noted, that the magnetic field jitter, which arises during the phase evolution, has to be extrapolated from the random walk model described in equation \ref{eqn:MagneticFieldJitter}, since all cycle times $T_\tn{cycle}>25\di{s}$ are larger than the studied phase evolution times $T_\tn{evol}\le12\di{s}.$\\ 
A fourth very tiny source of phase jitter is given once again by the initial thermal distribution of the modified cyclotron energy, which entails a jitter of the energy dependent systematic frequency shifts due to special relativity and the inhomogeneities of the electric and magnetic field. The corresponding phase jitters increase linearly with the evolution time. For a phase evolution time of $5\di{s},$ a modified cyclotron radius of $r_+^\tn{evol}=90(4)\;\mu\tN{m}$ and a conservative estimate of the axial temperature $T_z=5\di{K},$ all these phase jitters are smaller than $4^\circ.$ 
In comparison to the absolute value of the $5\di{s}$ phase jitter, these jitters can be neglected.

\section{Line-shape model of the $\Gamma$-resonance}\label{sec:lineShape}
The fundamental mechanism for a spin-flip transition is described by a Rabi oscillation, with the Rabi frequency $\Omega$, which scales linearly with the microwave amplitude. For a close-to-resonant drive of the Larmor frequency, $\omega_{L_0},$ with the frequency $\omega_\tn{MW}=\omega_{L_0}+\Delta,$ where $\Delta$ is an offset, the Rabi frequency is modified $\Omega' \equiv \sqrt{\Omega^2+\Delta^2}$ and the probability for a spin-flip is:
\begin{equation}
\label{eqn:RabiOffResonant}
p_\uparrow(t)=\frac{\Omega^2}{\Omega'^2} \sin^2\left(\frac{\Omega' t}{2}\right) .
\end{equation}
Some conditions of our experiment, e.g. the existing magnetic field fluctuations, prohibit coherent Rabi oscillations of the electron spin on the time-scale of a cyclotron frequency measurement. As a starting point for the construction of a line-shape model of the Larmor resonance we thus use the time averaged value of the squared sine term: 
\begin{equation}
\label{eqn:sfRatePure}
p_\tn {sf}=\frac{1}{2} \frac{\Omega^2}{\Omega^2+\Delta^2}.
\end{equation}
The spin-flip probability $p_\tn{sf}$ has a symmetric line-shape and saturates at a value of 0.5 for large Rabi frequencies. However, in the following we will exclusively focus on line-shapes well below saturation.\\
To take care of magnetic field drifts and fluctuations, we probe the spin-flip transition at the fixed frequency $\omega_\tn{MW}$ and simultaneously measure the modified cyclotron frequency, $\omega_+^\tn{PnA},$ which requires a line-shape model in dependence on the measured frequency ratios, $\Gamma^*=\omega_\tn{MW}/\omega_c:$
\begin{equation}
\label{eqn:sfRatePureGamma}
p_\tn {sf}(\Gamma^*)=\frac{1}{2} \frac{\Omega^2}{\Omega^2+\omega_{c_0}^2(\Gamma_0+\delta\Gamma_0-\Gamma^*)^2},
\end{equation}
where $\Gamma_0=\frac{\omega_{L_0}}{\omega_{c_0}}$ is the final value of interest and $\delta\Gamma_0$ are systematic shifts. For clarity reasons we ignore any systematic shifts for a moment. They will be discussed in detail in section~\ref{sec:systShifts}. Here, we focus exclusively on four independent frequency jitters, which modify the line-shape:\\
(1) The continuous thermalization of the axial mode during the $5\di{s}$ measurement process affects the cyclotron and the Larmor frequency due to the leading order inhomogeneity of the magnetic field, $B_2$. Since the measured modified cyclotron frequency is an average over the phase evolution time, only the Larmor frequency jitter alters the shape of the resonance, which causes the dominant asymmetric contribution of the line-shape ($E_z\tN{-}B_2$ asymmetry)~\cite{Verdu2004}:
\begin{flushleft}
\vspace{0.01cm}
\hspace*{-2cm}
\begin{minipage}{10cm}
\begin{equation}
\label{eqn:sfRatePure}
p_\tn{sf}(\Gamma^*)=\frac{1}{2} \int_0^\infty \frac{\Omega^2}{\Omega^2+\omega_{c_0}^2 \left(\Gamma_0+\alpha_\tn{B2}\cdot (E_z- \bar{E}_z)-\Gamma^* \right)^2}\frac{1}{k_B T_z}e^{-\frac{E_z}{k_B T_z}} \tN{d}E_z,\;\;\;\;\;\;\;\;\;\;\;\;\;\;
\end{equation}
\vspace{0.01cm}
\end{minipage}\\
\end{flushleft}
where $\alpha_\tn{B2} \equiv B_2\cdot (B_0 \cdot\omega_z^2 \cdot m_\tn{ion})^{-1}\cdot(\omega_{L_0}/\omega_{c_0}).$ Due to the same, but averaged effect in the modified cyclotron frequency, which is considered by the averaged term  $-\alpha_\tn{B2}\cdot \bar{E},$ the difference between the maximum and the mean value of the line-shape is small. For an axial temperature of $T_z=3.6\di{K}$ and a spin-flip probability of about $30\%$ the relative effect is:
\begin{equation}
\label{eqn:MaxMeanDeviation}
\frac{\Gamma_\tn{mean}-\Gamma_\tn{max}}{\Gamma_\tn{max}}\approx 3\cdot 10^{-13}.
\end{equation}
(2) Since the modified cyclotron frequency has Gaussian distributed thermal jitters and detection uncertainties, $\delta(\varphi_\tn{10ms})\approx 14(1)^\circ$, see figure~\ref{fig:DistributionsPnA}(a) and the 5th column of table~\ref{tab:10msPhaseJitter}, also the measured $\Gamma^*$ has the same relative jitter contribution, which has to be considered in the line-shape model by a Gaussian convolution.\\
(3) A further jitter of the measured $\Gamma^*$ values is caused by the random walk of the magnetic field during the measurement, see figure~\ref{fig:MagneticFieldJitter}(b). With a $5 \di{s}$ phase evolution time we have: $\delta \varphi_+^\tn{rw} = A\cdot 360^\circ \cdot \sqrt{5\di{s}}\cdot 5\di{s}/\sqrt{2}=10(1)^\circ,$ where: $A=0.00339(7)\di{s}^{-1.5}.$ In a good approximation this jitter can be convoluted in the same way as the thermal jitter and the detection uncertainty. The total Gaussian jitter given by the measurement of the modified cyclotron frequency is: 
\begin{equation}
\label{eqn:TotalJitterModCycFreq}
\delta \nu_+/\nu_+=\frac{\sqrt{7/6(\delta \varphi_{10ms})^2+(\delta \varphi_\tn{rw})^2}}{360^\circ \cdot \nu_+ \cdot T_\tn{evol}}\approx4(1)\cdot 10^{-10}.
\end{equation}
(4) A further jitter of the measured free cyclotron frequency is caused by the detection of the axial frequency. The fluctuation of the two dip measurements before and after the PnA-cycles is typically $30\di{mHz}.$ 
With an uncertainty of $15\di{mHz}$ for the averaged axial frequency, the final relative jitter of the free cyclotron frequency is: $1.7\cdot10^{-11},$ which can be neglected in the line-shape model.\\ 
In figure~\ref{fig:OneEmassResonance} the different contributions of the line-shape model are presented:
\begin{figure}
  \centering
  \includegraphics[width=0.8\textwidth]{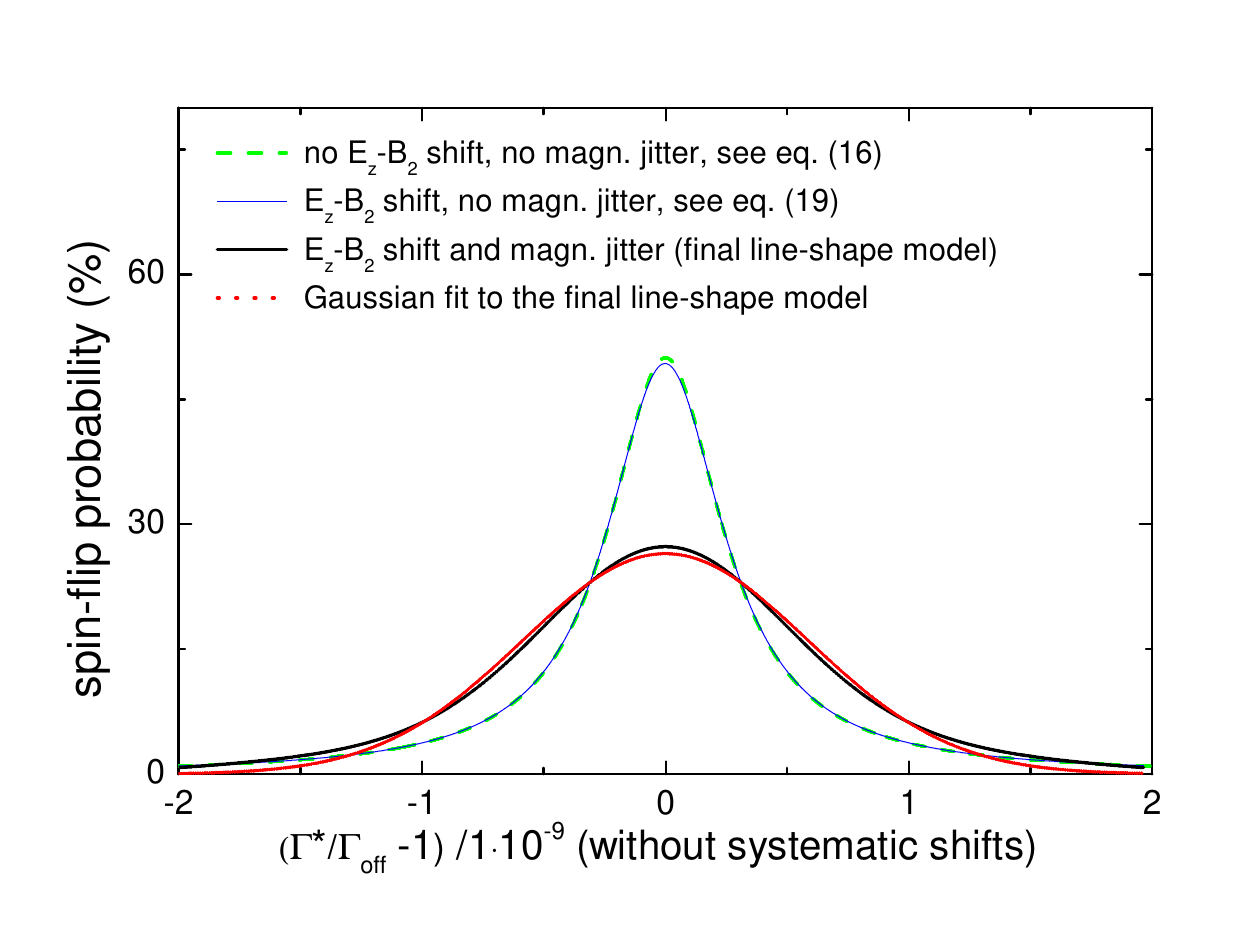}
  \caption{\label{fig:OneEmassResonance} Analysis of the different contributions to the line-shape model of the $\Gamma$-resonance: In green: The pure, averaged Rabi resonance is shown with a Rabi frequency of $\Omega=30\di{Hz}$ (see equation~(\ref{eqn:RabiOffResonant})). In blue: An additional $E_z\tN{-}B_2$ asymmetry with $T_z=3.6\di{K}$ (see equation~(\ref{eqn:sfRatePure})) is added. In black: The complete line-shape model is shown, also including the Gaussian phase jitter of $4\cdot 10^{-10}.$ In red: A Gaussian distribution is fitted to the complete line-shape model.}
\end{figure}Firstly, the pure averaged Rabi resonance with a Rabi frequency of $\Omega=30\di{Hz}$ (green line, equation~(\ref{eqn:RabiOffResonant})), secondly an additional $E_z\tN{-}B_2$ asymmetry with $T_z=3.6\di{K}$ (blue line, equation~(\ref{eqn:sfRatePure})), thirdly the complete line-shape model also including the Gaussian convoluted phase jitter of $\delta \Gamma^*/\Gamma=4\cdot 10^{-10}$ (black line). Finally the complete line-shape model is compared with a Gaussian distribution (red line). The deviation between the centroid of the line-shape model and the mean value of the Gaussian fit is only $1\cdot10^{-13}.$ This line-shape approximation is justified, as long as the maximal spin-flip probability is well below the saturation value of $50\%$. The characteristic model parameters, the maximal spin-flip probability of $\approx 27\%$ and the full width at half maximum (FWHM) of $\approx 1.3\cdot10^{-9}$ are in good agreement with the measured $\Gamma$-resonances, which have maximal spin-flip rates of $25\tN{-}45\%$ and FWHM's of $0.7\tN{-}1.35\cdot 10^{-9}$. 
The rather smaller resonance widths of the measured data might be explained by a partial cancellation of the magnetic field jitter in the measured frequency ratio $\Gamma^*$.\\
Finally the Gaussian line-shape, $\mathcal{G}(\tN{sf}_0,\mu,\sigma),$ is applied as a maximum-likelihood fit to the data~\cite{Sturm_PhD}:
\begin{eqnarray}
\label{eqn:ML-Fit}
\tN{log} \left[\mathcal{L}(\tN{sf}_0,\mu,\sigma)\right]=& \sum_{i=1}^{N_\tn{Sf}} \tN{log}(\mathcal{G}(\Gamma^\tn{*,Sf}(i);\tN{sf}_0,\mu,\sigma))+\nonumber\\
&\sum_{j=1}^{N_\tn{NonSf}} \tN{log}(1-\mathcal{G}(\Gamma^\tn{*,NonSf}(j);\tN{sf}_0,\mu,\sigma)),
\end{eqnarray} 
extracting the desired $\Gamma_\tn{res}$ value ($\mu$), the maximal spin-slip rate ($\tN{sf}_0$) and the width of the resonance ($\sigma$). A typical $\Gamma$-resonance has been shown in figure~\ref{fig:GammaReso}. Here the red line represents the maximum-likelihood fit, the dark gray area is the error band of the mean value, the black markers illustrate binned data and the bright gray area the binomial error of the binned data with the probability given by the maximum-likelihood fit.

\section{Statistical result}
\begin{figure}[htb]
  \centering
  \includegraphics[width=0.8\textwidth]{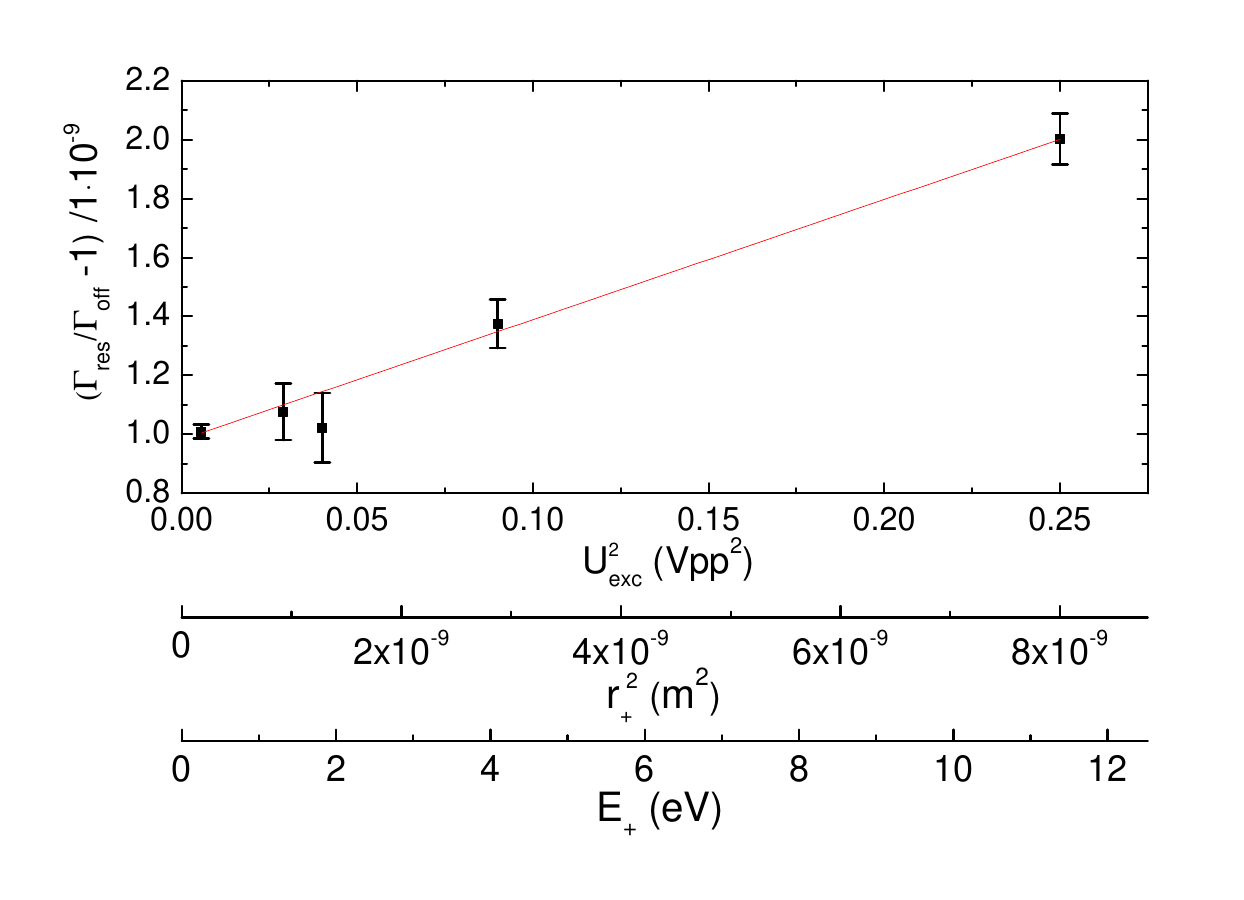}
  \caption{\label{fig:EmassResonances} Mean values from the Gaussian maximum-likelihood fits of the $\Gamma$-resonances at different $E_+$ during the PnA phase evolution times. A linear extrapolation to zero modified cyclotron energy is shown in red. The excited modified cyclotron radii and energies, which are indicated by the further two abscissae, can be calculated from the slope of the linear fit. For further details see text. ($\Gamma_\tn{off}=4376.210\;497\;791$, same scaling constant as in figure \ref{fig:GammaReso}.)}%
\end{figure}

Various $\Gamma$-resonances have been measured at different $E_+$  during the PnA phase evolution times, see figure~\ref{fig:EmassResonances}. The slope of the energy dependence is completely explained by a relativistic shift, see also section \ref{sec:specRel}: $m_{\frac{\delta \Gamma}{\Gamma}} \cdot U_\tn{exc}^2 \approx \delta \nu_+/\nu_+ \approx E_+/(m c^2)$, where $m_{\frac{\delta \Gamma}{\Gamma}}=4.07(35)\cdot 10^{-9} \di{Vpp}^{-2}$~\footnote{In that way, the excited modified cyclotron radius can be calculated as a function of the excitation amplitude of the first  $(10\di{ms})$ PnA pulse, denoted as $U_\tn{exc}:$ $r_+ \approx \sqrt{2 m_{\frac{\delta \Gamma}{\Gamma}} c^2/\omega_+^2}U_\tn{exc}=1.7889 \cdot 10^{-4} \di{m/Vpp} \cdot U_\tn{exc}.$}.
The extrapolated $\Gamma$ value at zero modified cyclotron energy is:
\begin{equation}
\label{eqn:GammaSatistics}
\Gamma_\tn{stat}=4376.210\;502\;112(102).
\end{equation}
So far no systematic corrections apart from the extrapolation in the modified cyclotron energy have been applied.

\section{Systematic shifts and uncertainties}\label{sec:systShifts}
\subsection{Electric field imperfections}\label{sec:elImper}
\begin{figure}
\begin{minipage}[hbt]{7.35cm}
\subfloat[]{\includegraphics[width=1.2\textwidth]{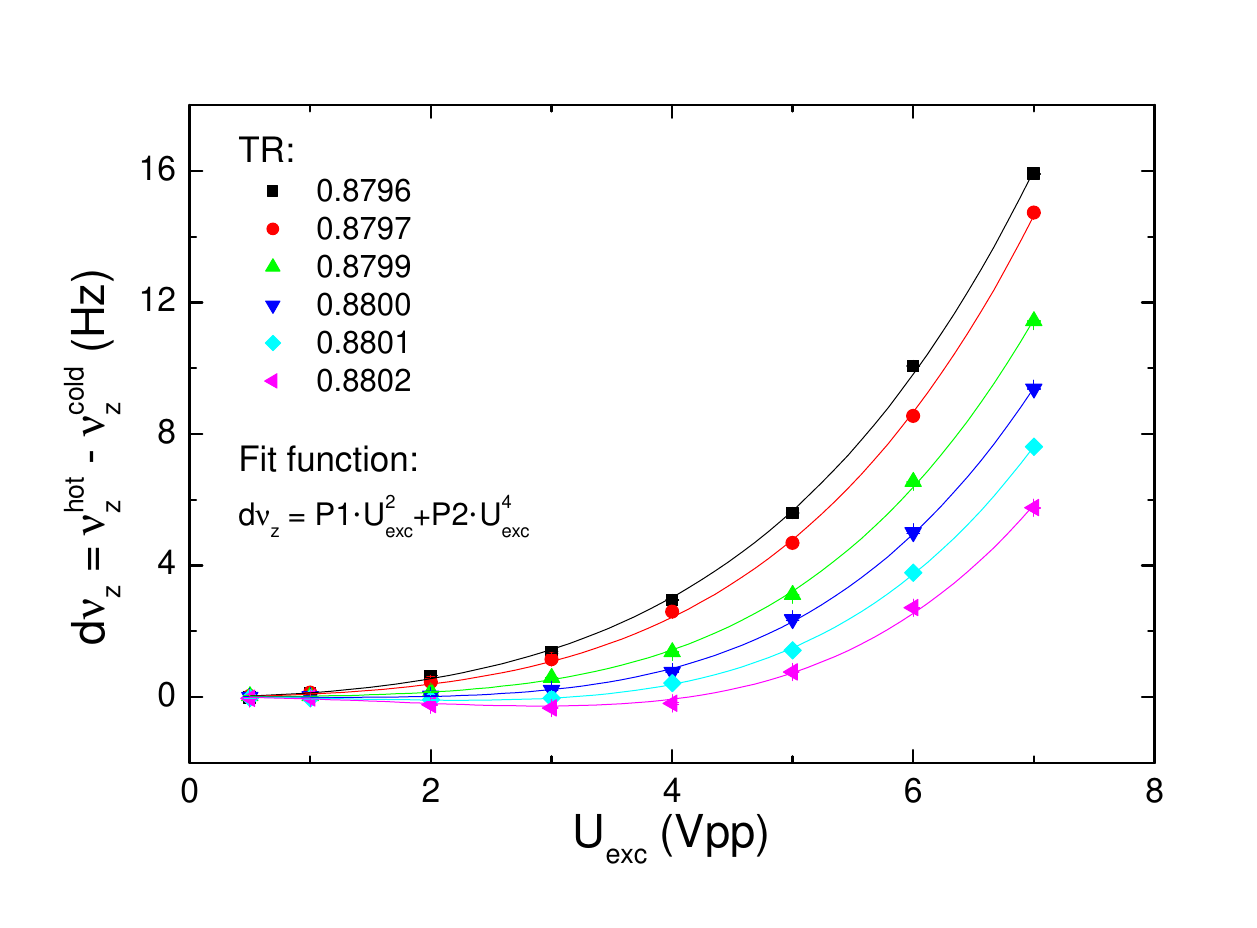}}
\end{minipage}
\hfill
\begin{minipage}[hbt]{7.35cm}
\subfloat[]{\includegraphics[width=1\textwidth]{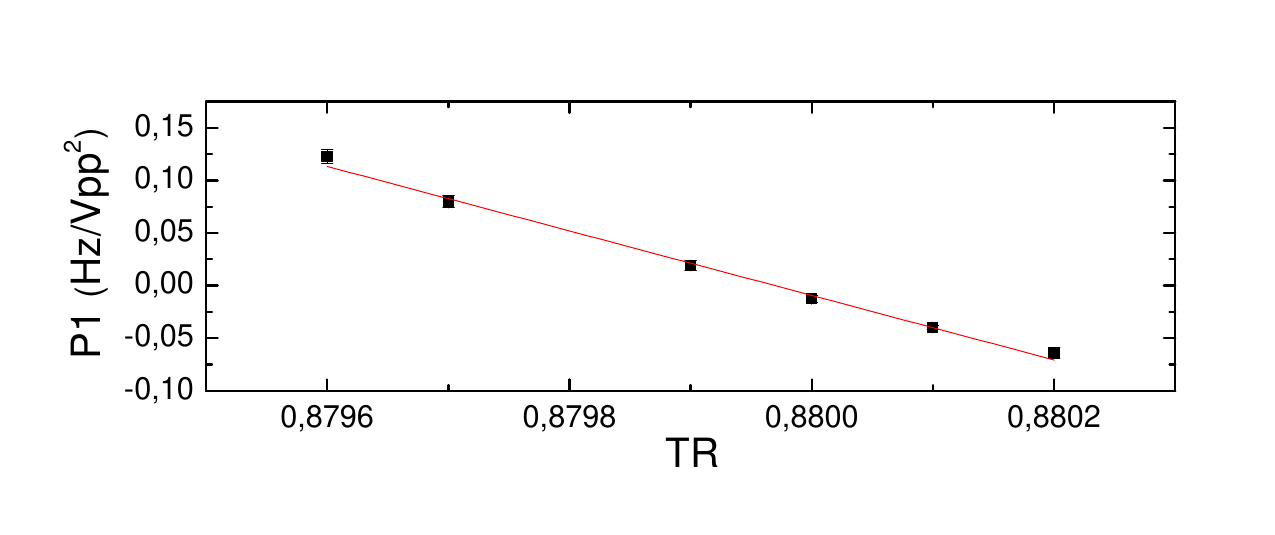}}\\
\subfloat[]{\includegraphics[width=1\textwidth]{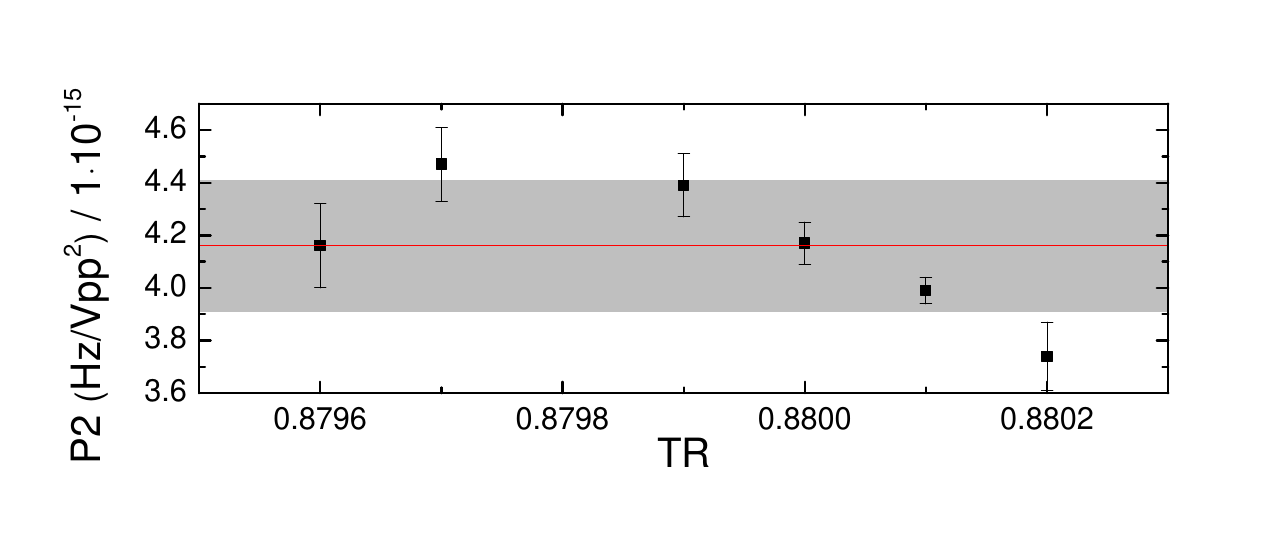}}
\end{minipage}
  \caption{\label{fig:TROpti}TR optimization in the PT via magnetron burst excitation. (a): Axial frequency differences between excited and thermalized magnetron modes in dependence on the magnetron burst amplitude with a fixed pulse length of $10\di{ms}$ for different TR's. (b): $\tN{P1}$ ($\propto C_4$), extracted from a two parameter fit as a function of the TR. (c): $\tN{P2}$ ($\propto C_6$) as a function of the TR. For details see text.}  
\end{figure}

Deviations from the ideal electric quadrupole potential, see equation~(\ref{eqn:Potential:zaxis}), lead to energy dependent eigenfrequency shifts \cite{Brown1986_Geonium,Ketter20141}. In the optimal design of a five electrode configuration with grounded endcaps, the Penning trap is \textit{orthogonal}, so that the axial frequency, which is proportional to $\sqrt{C_2},$ does not depend on the tuning ratio (TR), $\tN{TR} \equiv U_\tn{c}/U_\tn{r}.$ Moreover, the electric potential of an ideal five electrode Penning trap is doubly \textit{compensated}: $C_4=C_6=0$~\cite{Gabrielse_1989Traps}. 
Due to small imperfections of our trap design only $C_4$ can be completely nulled, while a tiny $C_6$ remains. During each ion creation process, regions on the electrode surfaces that are electrically isolating, e.g. due to frozen rest gas, can trap the unwanted ion species. These so-called patch potentials slightly modify the trapping potential in the PT, demanding a TR optimization right before the automatized measurement process starts. For this purpose axial frequency shifts, generated by magnetron burst excitations, are studied in dependence of the TR for different burst amplitudes, see figure~\ref{fig:TROpti}(a).
For a two parameter fit of the axial frequency shifts in dependence of the magnetron burst amplitude only the leading order $C_4$ and $C_6$ terms are considered:
\begin{eqnarray}
\label{eqn:TRFit}
\tN{d}\nu_z &= -\frac{3}{2}\frac{1}{d_\tn{char}^2}\frac{C_4}{C_2}\nu_z \cdot\eta^2\cdot U_ \tn{exc}^2+2.8125\frac{1}{d_\tn{char}^4}\frac{C_6}{C_2}\nu_z \cdot\eta^4\cdot U_ \tn{exc}^4\\
&= \tN{P1}\cdot U_ \tn{exc}^2 + \tN{P2}\cdot U_ \tn{exc}^4.
\end{eqnarray}
The magnetron radius is proportional to the magnetron burst amplitude: $\rho_-=\eta \cdot U_\tn{exc}.$ In figure~\ref{fig:TROpti}(b) P1 is plotted against the TR. The optimal TR at $\tN{P1}=C_4=0$ is extracted from a linear fit: $\tN{TR}_\tn{opt}=0.8799693(51)$. At large magnetron radii additional higher-order terms of the electric potential have to be considered in the fitting routine. For this reason a systematic uncertainty of $1\cdot10^{-5}$ is added to the optimized TR. During the complete measurement period of 4.5 months the optimal TR has only shifted by $<1\cdot10^{-5}$, so that we assume a conservative estimation for the relative uncertainty of the optimal TR of $2\cdot10^{-5}.$ $C_4$ can be split in a TR-independent and a TR-dependent part: $C_4=E_4+\tN{TR} \cdot D_4.$
Relying on a calculated value for $D_4^\tn{calc}=-0.616,$ we estimate the uncertainty of $C_4$: 
\begin{equation}
\label{eqn:C4Error}
\delta C_4=\left|D_4^\tn{calc}\right| \cdot \tN{TR} \cdot 2\cdot 10^{-5} =1.1\cdot10^{-5}.
\end{equation}
The eigenfrequency shifts described by the $C_4$-matrix~\cite{Brown1986_Geonium}:
\begin{flushleft}
\vspace{0.01cm}
\hspace*{-2.4cm}
\begin{minipage}{10cm}
\begin{equation}
\label{eqn:C4}
\left( \begin{array}{c}
\Delta \nu_+ / \nu_+\\
\Delta \nu_z / \nu_z\\
\Delta \nu_- / \nu_-\end{array} \right) = 
\frac{6 C_4 }{q_\tn{ion} U_r C_2^2}
\left( \begin{array}{ccc}
\frac{1}{4}(\nu_z/\nu_+)^4 & -\frac{1}{2}(\nu_z/\nu_+)^2  & -(\nu_z/\nu_+)^2  \\
-\frac{1}{2}(\nu_z/\nu_+)^2 & \frac{1}{4} & 1 \\
-(\nu_z/\nu_+)^2 & 1 & 1\end{array} \right)
\cdot
\left( \begin{array}{c}
E_+\\
E_z\\
E_-\end{array} \right)\;\;\;
\end{equation}
\vspace{0.01cm}
\end{minipage}\\
\end{flushleft}
cause a relative uncertainty of $\Gamma:$ 
\begin{equation}
\label{eqn:C4}
\left(\frac{\delta \Gamma}{\Gamma}\right)_{C_4}=5.0 \cdot 10^{-13}.
\end{equation}
An estimation of $C_6$ is possible by the fitted P2 value, see figure~\ref{fig:TROpti}(c) and the calibration of the magnetron radius. With the slope $a=-307(10) \di{Hz Vpp}^{-2}$ in figure~\ref{fig:TROpti}(b) and the predicted value of $D_4$, the proportional constant is: 
\begin{equation}
\label{eqn:eta}
\eta=\sqrt{- \frac{2 a \cdot C_2\cdot \tN{d}_\tn{char}^2}{3 \nu_z D_4}}=5.1(1)\cdot 10^{-5},
\end{equation}
where $C_2=-0.5504,$ see equation~(\ref{eq:axialFreq}). The estimation of $C_6$:
\begin{equation}
\label{eqn:C4Splitting}
C_6=\frac{\tN{P2}\cdot C_4 \cdot d_\tn{char}^4}{2.8125 \cdot \nu_z \cdot \eta^4}=-0.016(1)
\end{equation}
 is of the same size as the predicted value $C_6^\tn{calc}=-0.012.$ Assuming a conservative $C_6$-uncertainty of $100\%,$ the relative uncertainty of $\Gamma$ is:
\begin{equation}
\label{eqn:C6}
\left(\frac{\delta \Gamma}{\Gamma}\right)_{C_6}=5.9 \cdot 10^{-14},
\end{equation}
 calculated with the formulas given in~\cite{Ketter20141}.

\subsection{Magnetic field imperfections}
The large magnetic bottle in the AT generates a residual magnetic inhomogeneity in the PT. 
The linear gradient of the magnetic field, $B_1=-13.41(23)\cdot 10^{-3}\di{T/m}$~\cite{Sturm_PhD}, shifts the center of the ion motion in axial direction due to the force acting on the magnetic moment generated by the cyclotron motion. At the maximally applied modified cyclotron radius of $r_+=90\; \mu\tN{m}$ the center of the motion shifts $\Delta z = -21\di{nm},$ causing the same relative shifts of the free cyclotron frequency and the Larmor frequency of $-7.4\cdot 10^{-11},$ which completely cancel in the frequency ratio $\Gamma.$\\
The residual second order magnetic field inhomogeneity in the PT, $B_2=1.01(0.20) \di{T/m}^2$, see also table~\ref{tab:B2measurementsEmass}, evokes energy dependent shifts of both the eigenfrequencies and the Larmor frequency~\cite{Brown1986_Geonium}:
\begin{equation}
\label{eqn:B2}
\left( \begin{array}{c}
\Delta \nu_+ / \nu_+\\
\Delta \nu_z / \nu_z\\
\Delta \nu_- / \nu_-\\
\Delta \nu_L / \nu_L 
\end{array} \right)
=
\frac{B_2 }{B_0 m_\tn{ion} \omega_z^2}
\left( \begin{array}{ccc}
-(\nu_z/\nu_+)^2 & 1  & 2  \\
1 & 0 & -1 \\
2 & -1 & -2\\
-(\nu_z/\nu_+)^2 & 1  & 2
\end{array}\right)
\cdot
\left( \begin{array}{c}
E_+\\
E_z\\
E_-\end{array} \right).
\end{equation}
Since the Larmor frequency and the modified cyclotron frequency shift by the same amount, the magnetic shift cancels to a large extent in the frequency ratio $\Gamma$:
\begin{equation}
\label{eqn:B2}
\left(\frac{\Gamma_\tn{stat}-\Gamma_\tn{off}}{\Gamma_\tn{off}}\right)_{B_2}=-1.36(26) \cdot 10^{-12}.
\end{equation}
The error is dominated by the uncertainty of $B_2$.

\subsection{Special relativity}\label{sec:specRel}
The relativistic increase of the ion mass accounts for a relative shift of the cyclotron frequency:
\begin{equation}
\label{eqn:SpecialRel}
\frac{\Delta \nu_c}{\nu_c} =\frac{1}{\gamma}-1,
\end{equation}
where $\gamma \equiv (1-v^2/c^2)^{-1/2}$ is the relativistic Lorentz factor and $v$ is the velocity of the ion. At the same time the Larmor frequency is shifted by the Thomas precession $(1-\gamma)\nu_c$~\cite{Sturm_PhD}. In contrast to the \gf measurement of the free electron, the relative shift of the Larmor frequency is suppressed by a factor of $\nu_c/\nu_L$ with respect to the relative shift of the cyclotron frequency. For this reason the slope of the $\Gamma$-resonances at different modified cyclotron energies, see figure~\ref{fig:EmassResonances}, is dominated by the relativistic mass increase:
\begin{equation}
\label{eqn:RelativisticSlope}
(\delta \Gamma/\Gamma)_\tn{relat}\approx \delta \nu_c/\nu_c\approx \delta \nu_+/\nu_+\approx E_+/(m_\tn{ion}c^2)\approx8.9\cdot10^{-11} \di{eV}^{-1}\cdot E_+.
\end{equation}
After the extrapolation to zero modified cyclotron energy, the speed of the ion is mainly given by the axial mode, $\bar{v}_z=87(2) \di{m/s}$ at an axial temperature $T_z=5.44(22)\di{K},$ resulting in a relativistic shift of $\Gamma:$
 \begin{equation}
\label{eqn:Relat}
\left( \frac{\Gamma_\tn{stat}-\Gamma_\tn{off}}{\Gamma_\tn{off}}\right)_\tn{relat}=4.20(17) \cdot 10^{-14}.
\end{equation}
A further relativistic shift of the Larmor frequency 
is generated by the additional motional magnetic field: $\Delta B=\gamma/c^2 \;(\vec{v}\times \vec{E}).$ Even for the $\Gamma$-resonance with the largest modified cyclotron energy ($r_+=90\;\mu\tN{m}$) and the corresponding radial electric field $E_r=19 \di{V/m}$ the relative magnetic field shift is only: $\Delta B/B=8\cdot10^{-13}.$

\subsection{Image current shift}
In presence of the axial resonator with impedance  $Z_\tn{LC}(\omega_z),$ the following drag force $F_z^\tn{drag}(\omega_z)$ acts on the axial motion of the ion~\cite{Sturm_PhD}:  
\begin{equation}
\label{eqn:ImageCurrentShift}
F_z^\tn{drag}(\omega_z)=-\frac{q_\tn{ion}^2}{D_\tn{eff}^2}Z_\tn{LC}(\omega_z) \cdot \dot{z} \equiv -2 \gamma m_\tn{ion} \cdot \dot{z},
\end{equation}
Consequently, the damped axial motion can be described by: $z(t)=z_0 \exp\left[- (\gamma -i\sqrt{\omega_\tn{z,ideal}^2-\gamma^2}) t\right].$ Since the resonator impedance has a complex value, the imaginary part of the damping constant $\gamma$ causes a particular frequency shift, the \textit{image current shift}: $\Delta\omega_z=\omega_z^\tn{meas.}-\omega_z^\tn{ideal}=-\tN{Im}(\gamma).$ For the determination of the axial frequency via the dip-signal this systematic shift is already included in the line-shape model, see equation~(\ref{eqn:DipFit}). However, the image current shift of the modified cyclotron frequency caused by the imaginary part of the impedance of the cyclotron resonator has to be corrected:
\begin{equation}
\label{eqn:imageCurrentShiftCyclotron}
\frac{\Delta\omega_+}{\omega_+}\approx-\frac{\tN{Im}(\gamma)}{\omega_+}=-\frac{q_\tn{ion} \tN{Im}(Z_\tn{LC}(\omega_+))}{2 m_\tn{ion} \omega_+ D_\tn{rad}^2}.
\end{equation}
This resonator is used for ion identification directly after the creation process and is shifted away by a varactor-diode during the measurement process.
\begin{figure}[htb]
  \centering
	\includegraphics[width=0.7\textwidth]{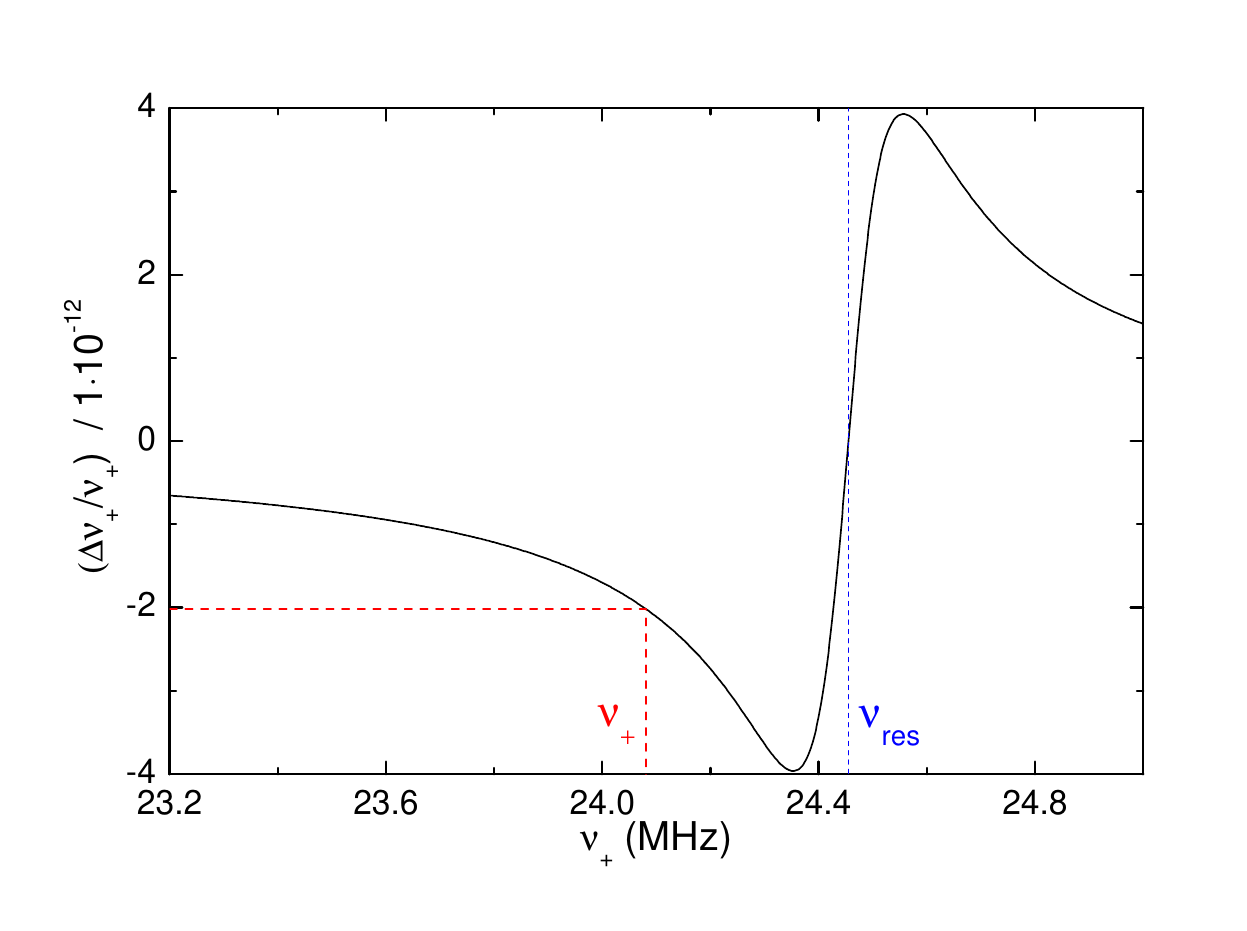}	
	\caption{\label{fig:ImageCurrentCyclotronEmass}The image current shift caused by the cyclotron resonator in the PT as a function of $\nu_+.$ The resonator parameters are listed in table~\ref{tab:CyclotronReso}.}
\end{figure}Figure~\ref{fig:ImageCurrentCyclotronEmass} illustrates the image current shift in dependence of the modified cyclotron frequency. The corresponding systematic shift of $\Gamma$ is:
\begin{equation}
\label{eqn:ImageCurrentShift}
\left(\frac{\Gamma_\tn{stat}-\Gamma_\tn{off}}{\Gamma_\tn{off}}\right)_\tn{image current}=2.20(55) \cdot 10^{-12}.
\end{equation}

\subsection{Image charge shift}
The induced charges at the electrode surfaces generate an additional effective electric potential, which shifts the radial eigenfrequencies of the ion. An analytical calculation of this \textit{image charge shift} 
is given in~\cite{Sturm2013_ImageChargeShift}:
\begin{equation}
\label{eqn:ImageChargeShiftI}
\frac{\Delta \nu_c}{\nu_c}=1.92\frac{m_\tn{ion}}{8 \pi \epsilon_0 r^3 B_0^2},
\end{equation}
where the corresponding relative uncertainty has been estimated to $5\%.$ It should be noted that the shift scales with the inverse cubic trap radius, providing in the future a possible significant reduction of this effect by using a larger trap. The image charge shift is by far the largest systematic shift of $\Gamma$ and dominates the systematic uncertainty budget: 
\begin{equation}
\label{eqn:ImageChargeShiftII}
\left( \frac{\Gamma_\tn{stat}-\Gamma_\tn{off}}{\Gamma_\tn{off}} \right)_\tn{image charge}=2.824(141) \cdot 10^{-10}.
\end{equation}

\subsection{Modified cyclotron frequency - PnA method} 
The detection of the modified cyclotron frequency has already been discussed at length in section~\ref{sec:PnA}. The uncertainties due to magnetic field fluctuations and the technical readout uncertainty are already included in the statistical error. Here a more subtle effect is studied, which occurs during the second pulse.
\subsubsection{Dipole contribution of the second PnA pulse}~\\
The so-called \textit{quadrupole excitation}, where the exciting field is connected to one half of the split correction electrode, features also small dipole components in radial and axial directions. These components have been numerically determined based on a finite element simulation using \Comsol ~\cite{comsol}.
\begin{equation}
\label{eqn:RealQuadExc}
\vec{F}_\tn{quad. real}=q_\tn{ion} 
\left( \begin{array}{c}
64.7 + 38410.1 \di{m}^{-1}\cdot z\\
0\\
75.7+38410.1 \di{m}^{-1}\cdot x\end{array} \right)
A_\tn{quad.} \sin(\omega' t+\varphi_0).
\end{equation}
During the second PnA pulse the resonant quadrupole excitation at the sideband frequency $\nu_+ + \nu_z$ competes with an off-resonant dipole excitation at $\nu_+$. As a result, the read-out phase has a systematic shift dependent on the modified cyclotron phase with respect to the phase of the coupling pulse. In the worst-case scenario of an in-phase excitation, starting amplitudes of $z=r_+=15~\mu\tN{m}$ and a final amplitude of $z=100~\mu\tN{m}$ the numerically calculated shift of the modified cyclotron frequency is smaller than $0.05\di{mHz}$ for a $10\di{ms}$ pulse length and a phase evolution time of $5\di{s}$. The corresponding relative uncertainty of the frequency ratio is smaller than 
\begin{equation}
\label{eqn:2ndPnAPulseDipolar}
\left( \frac{\delta\Gamma}{\Gamma}\right)_\tn{dipole contr} < 3 \cdot 10^{-12}.
\end{equation} 

\subsection{Axial frequency - line-shape model of the dip-signal}
The line-shape model of the axial dip-signal, see figure~\ref{fig:AxialResonatorAndFit}(b), is introduced in~\cite{Sturm_PhD}:
\begin{eqnarray}
\label{eqn:DipFit}
u_\tn{n}^\tn{Dip}&(\omega,|\omega_z,A,\tau,Q,\omega_\tn{res},u_\tn{en},\kappa_\tn{det}) \\
&=10 \log_{10} \left[ A \cdot \tN{Re}(Z_\tn{tot}(\omega|\omega_z,\tau,Q,\omega_\tn{res},R_p)/R_p)+u_\tn{en}^2 \right]+\kappa_\tn{det} \cdot (\omega-\omega_\tn{res}),\nonumber
\end{eqnarray}
where the real part of the resonance is:
\begin{eqnarray}
\label{eqn:realtotalImpedance}
\tN{Re}(Z_\tn{tot})&=\frac{R_p\left(\omega_\tn{res}\omega (\omega^2-\omega_z^2)\right)^2}{\left(\omega_{res}\omega (\omega^2-\omega_z^2) \right)^2+ \left(Q (\omega^2+\omega_\tn{res}^2)(\omega^2-\omega_z^2)  - \omega_\tn{res} \omega^2/ \tau \right)^2}.
\end{eqnarray}
From the fit of the dip spectrum the axial frequency of the ion, $\omega_z,$ the amplification amplitude, $A,$ and the cooling time constant, $\tau,$ are extracted each time. The line-shape model of the dip requires four input parameters: (1) the resonance frequency of the resonator, $\nu_\tn{res},$ (2) the quality factor of the resonator, $Q,$ (3) the noise contribution of the cryogenic amplifier, $u_\tn{en},$ and (4) a frequency dependence of the detection system, $\kappa_\tn{det},$ which is of first order in the logarithmic scale of the noise spectrum. All four parameters are derived from fits of the complete resonator spectrum, see figure~\ref{fig:AxialResonatorAndFit}(a). Their uncertainties account for systematic uncertainties of the axial frequency, see table~\ref{tab:DependenceVzOfResParam}.
\begin{table}
\centering
\caption{\label{tab:DependenceVzOfResParam}Input parameters for the line-shape fit of the axial dip-signal, extracted from axial resonator fits. From the uncertainties of these parameters (second column) and their impact on the axial frequency (third column) systematic uncertainties of the axial frequency are calculated (fourth column).}
\footnotesize
\begin{tabular}{l c c c}
\br
resonator  & value & axial dependence & uncertainty in \\
parameter & &  $d \nu_z / d (\tN{par})$ &  $\nu_z \di{(mHz)}$\\
\mr
$\nu_{res}$ 	& $670\;890(8)\di{Hz}$ 			& $-1.8(5) \cdot 10^{-4}$  			& $1.4(4)$\\
Q 		& $670(40)$ 					& $1.9(8) \cdot 10^{-5} \di{Hz}^{-1}$ 		& $0.8(2)$\\
$\kappa_\tn{det}$		& $2.5(8.0)\cdot10^{-5} \di{dBVrms/Hz}$ & $-0.6(2.2) \di{Hz}^2\tN{/dBVrms}$ 	 & $0(1)$\\
$u_\tn{en}$ 		& $1.2(2)\di{Vrms}$ 			& $0(2)\cdot 10^{-2} \di{Hz/Vrms}$ 		& $0(4)$\\
\br
\end{tabular} 
\end{table}
\normalsize
With a conservative estimate of the total error for the dip-signal line-shape model of $4.5\di{mHz},$ the relative uncertainty of the cyclotron frequency and thus the frequency ratio is:
\begin{equation}
\label{eqn:DipSystError}
\left( \frac{\delta \Gamma}{\Gamma}\right)_\tn{line-shape axial}=\frac{\delta \nu_c}{\nu_c}=\frac{\nu_z \; \delta \nu_z}{\nu_c^2}= 5.2\cdot10^{-12}.
\end{equation}
Since the axial frequency is measured before and after the simultaneous PnA measurement of the modified cyclotron frequency and the probing of the Larmor frequency, the drift between the two axial frequency measurements has to be considered. The mean value of the frequency differences of the two dip measurement is $|\tN{d}\bar{\nu}_z|=2.5\di{mHz}.$ 
Considering that basically only the last of the ten subsequent PnA-cycles determines the measured frequency ratio, $\Gamma^*,$ the second axial dip measurement is temporally closer to the relevant PnA-cycle, so that due to the axial drift we assume a relative uncertainty of $\Gamma$ of $\ll 1.2\cdot10^{-12}.$

\subsection{Drift of the magnetron frequency}
During the $4.5$ months of data taking the magnetron frequency has been measured via the double-dip method only three times. The largest measured shift of $\Delta \nu_-=0.2\di{Hz}$ corresponds to the following uncertainty of $\Gamma:$
\begin{equation}
\label{eqn:MagnetronUncert}
\left( \frac{\delta\Gamma}{\Gamma}\right)_{\delta \nu_-}=3.22 \cdot 10^{-12}.
\end{equation}

\subsection{Consistency checks in the PT}\label{sec:B2inPT}
\begin{figure}
\subfloat[]{\includegraphics[width=0.49\textwidth]{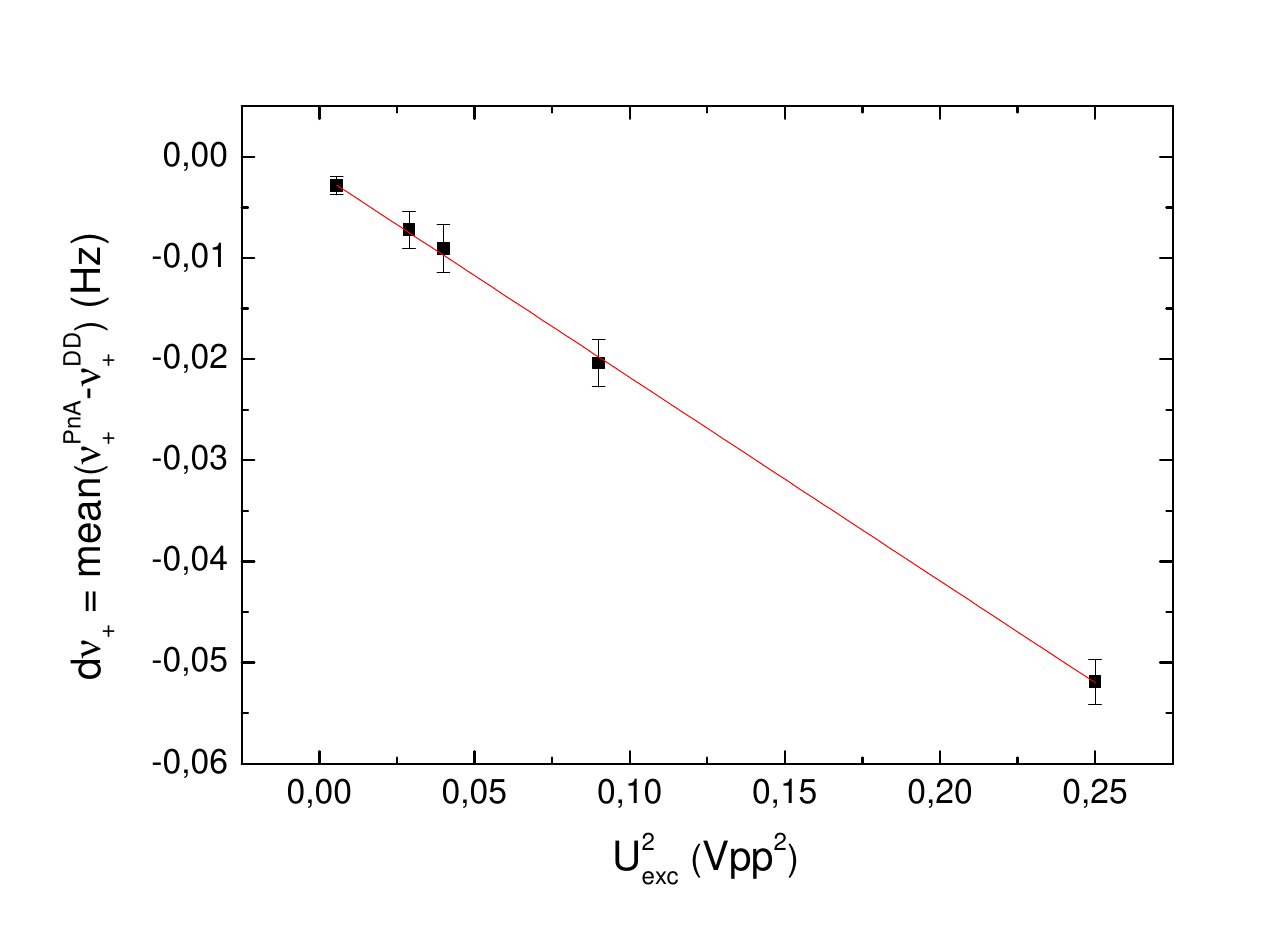}}\hfill
\subfloat[]{\includegraphics[width=0.49\textwidth]{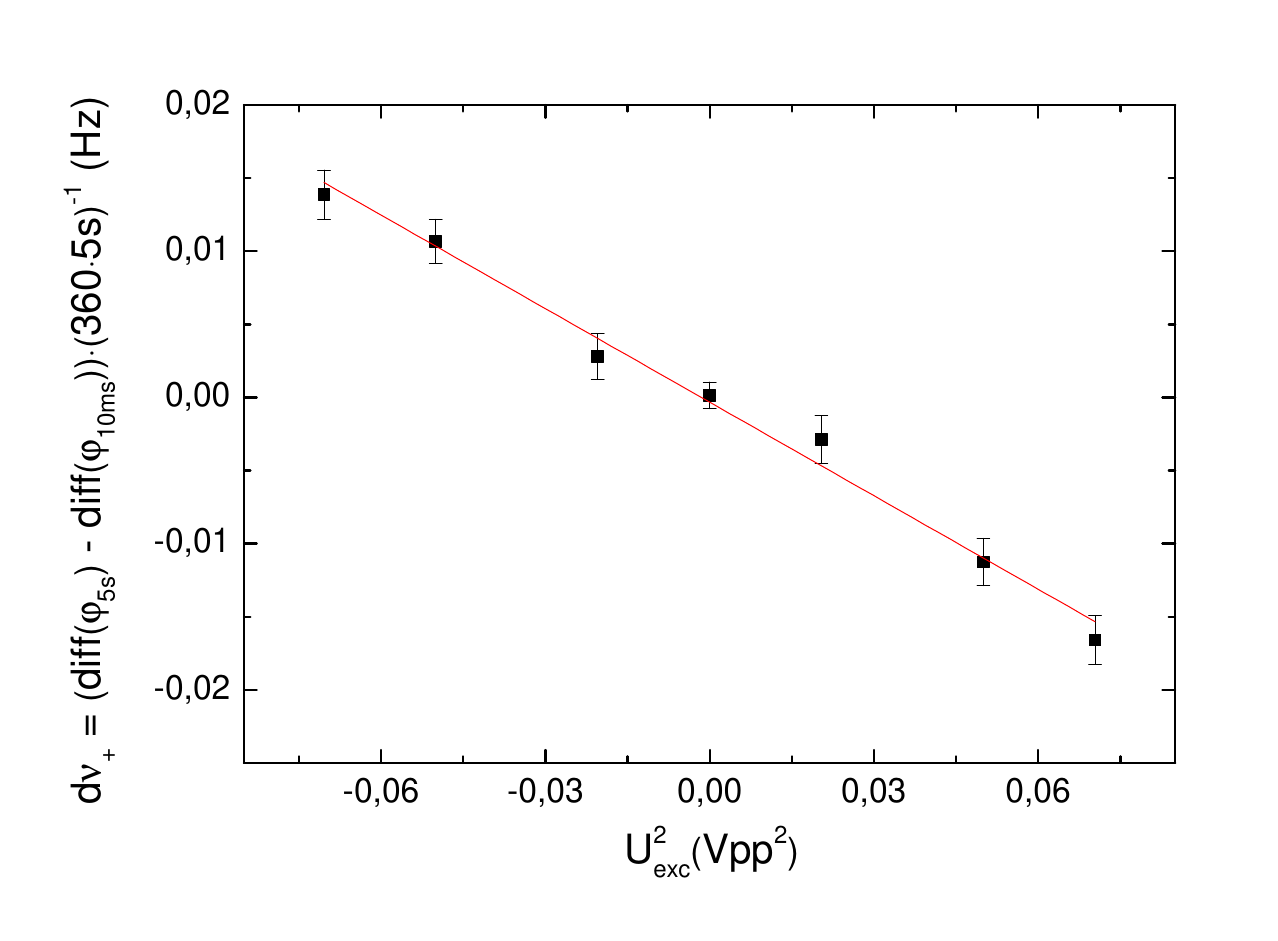}}\hfill
\subfloat[]{\includegraphics[width=0.49\textwidth]{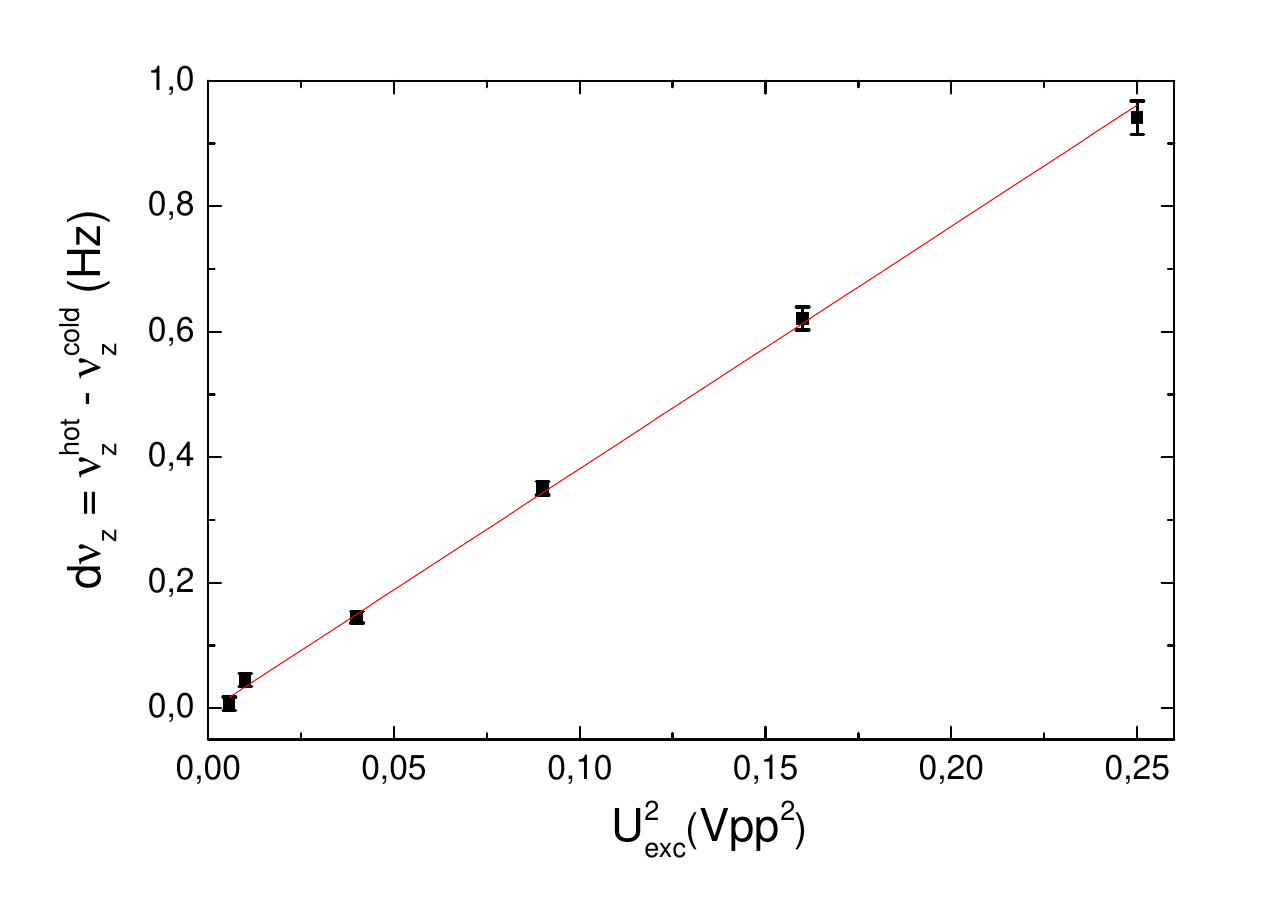}}\hfill
\subfloat[]{\includegraphics[width=0.49\textwidth]{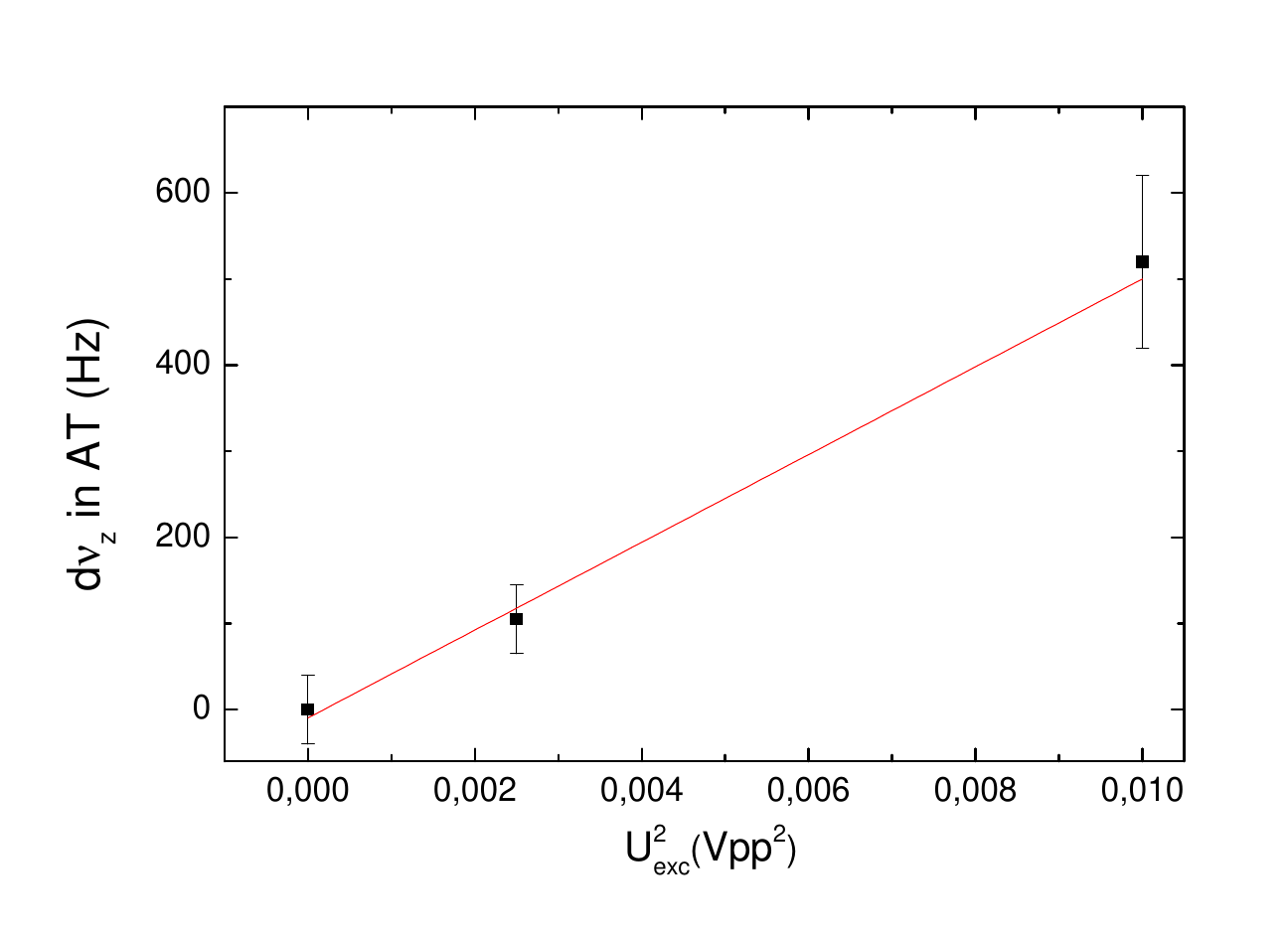}}
  \caption{\label{fig:consistCheck}Consistency checks: Various eigenfrequency shifts in dependence of the squared amplitude of the first PnA pulse, which is proportional to the modified cyclotron energy. Detailed descriptions are given in the text.} 
\end{figure}

At different energies of the modified cyclotron mode, which are proportional to the squared motional radius and thus to the squared amplitude of the first PnA excitation pulse ($U_\tn{exc}$), all three eigenfrequencies shift due to the relativistic mass increase and the magnetic inhomogeneity in the PT. From the combination of the measured four eigenfrequency shifts the magnetic inhomogeneity $B_2$ can be calculated in different, independent ways, providing a consistency check of the energy dependent systematic shifts:\\
(1) The slope $m_{\frac{\delta \Gamma}{\Gamma}}=4.07(35)\cdot 10^{-9} \di{Vpp}^{-2}$ of the different $\Gamma$-resonances with respect to the different modified cyclotron energies, see figure~\ref{fig:EmassResonances}, is mainly given by the relativistic shift:
\begin{equation}
\label{eqn:SlopeFinalResos}
m_{\frac{\delta \Gamma}{\Gamma}} \cdot U_\tn{exc}^2 \approx -\frac{\delta \nu_+}{\nu_+} \approx\frac{E_+}{m c^2}.
\end{equation}
(2) The differences of the modified cyclotron frequencies, determined in each measurement cycle, firstly by the double-dip and secondly by the PnA method, are shown in figure~\ref{fig:consistCheck}(a)
as function of the different modified cyclotron energies, which are represented here and in the following plots by the squared excitation amplitude $U_\tn{exc}^2$  of the first PnA pulse. These two different techniques are in remarkable agreement at zero modified cyclotron energy: $<\nu_+^\tn{PnA}-\nu_+^\tn{DD}>/\nu_+=7(10)\cdot10^{-11}$\footnote{Error includes uncertainty of the line-shape model for the dip.}.
The slope  $m_{\nu_+^\tn{PnA}-\nu_+^\tn{DD}}=-0.201(10) \di{Vpp}^{-2}$  is given by the relativistic- and the $B_2$-shift:
\begin{equation}
\label{eqn:SlopevpPnAMinusvpDD}
m_{\nu_+^\tn{PnA}-\nu_+^\tn{DD}} \cdot U_\tn{exc}^2 \approx -\frac{E_+\cdot \nu_+}{m \cdot c^2} - \frac{B_2 \cdot E_+}{B_0 \cdot m \cdot (2\pi)^2 \cdot \nu_+}.
\end{equation}
The same frequency shift has been measured in subsequent PnA measurements at randomly chosen modified cyclotron energies ($U_\tn{exc}=0.14,\;0.2,\;0.3 \di{Vpp}$). In figure~\ref{fig:consistCheck}(b) the phase differences are plotted against the energy differences. The linear fit  has the following slope: $m_{\nu_+ \tn{by PnA}}=-0.213(13) \di{Hz}\cdot \tN{Vpp}^{-2}.$\\
(3) Also the axial frequency shift, caused by the different modified cyclotron energies and the axial $B_2$-shift has been detected, see figure~\ref{fig:consistCheck}(c). The slope $m_{\nu_z}=3.863 (84)\di{Hz}\cdot \tN{Vpp}^{-2}$ depends quasi exclusively on the $B_2$-shift at different cyclotron energies:
\begin{equation}
\label{eqn:SlopevzInPT}
m_{\nu_z} \cdot U_\tn{exc}^2 \approx \frac{B_2 \cdot E_+}{B_0 \cdot m \cdot (2\pi)^2 \cdot \nu_z}.
\end{equation}
(4) The axial frequency shift in the AT with the slope $m_{\nu_z}^\tn{AT}= 50.9(1.1)\cdot 10^3 \di{Hz}\cdot \tN{Vpp}^{-2},$ see figure~\ref{fig:consistCheck}(d), depends on the magnetic bottle in the AT:
\begin{equation}
\label{eqn:SlopevpInAT}
m_{\nu_z^\tn{AT}} \cdot U_\tn{exc}^2 \approx \frac{B_2^\tn{AT} \cdot E_+}{B_0^\tn{AT} \cdot m \cdot (2\pi)^2 \cdot \nu_z^\tn{AT}}.
\end{equation}
In table~\ref{tab:B2measurementsEmass} the different values for the $B_2$ in the PT are summarized. They all agree within their error bars and show no sign for further unknown energy-dependent shifts, validating the model of systematics discussed above. 
\begin{table}
\centering
\caption{\label{tab:B2measurementsEmass}Consistency checks: Various approaches have been used to determine $B_2$ in the PT.}
\footnotesize
\begin{tabular}{l c }
\br
combination of equations: & $B_2$ ($T/m^2$)\\
\mr
(\ref{eqn:SlopeFinalResos}) and (\ref{eqn:SlopevpPnAMinusvpDD}) ($m_{\nu_+^\tn{PnA}-\nu_+^\tn{DD}}$)& $1.01(20)$\\
(\ref{eqn:SlopeFinalResos}) and  (\ref{eqn:SlopevpPnAMinusvpDD}) ($m_{\nu_+ \tn{by PnA}}$) & $1.12(22)$\\
(\ref{eqn:SlopeFinalResos}) and  (\ref{eqn:SlopevzInPT})  & $1.053(93)$\\
(\ref{eqn:SlopevpPnAMinusvpDD}) and (\ref{eqn:SlopevpInAT})  & $1.40(50)$\\
\br
\end{tabular} 
\end{table}
\normalsize

\section{Result} 
Extrapolating the various measured frequency ratios to zero modified cyclotron energy, our final measured frequency ratio is:
\begin{equation}
\label{eqn:GammaSatisticsII}
\Gamma_\tn{stat}=4376.210\;502\;112(102),
\end{equation}
see also equation (\ref{eqn:GammaSatistics}). If we correct this value by the systematic shifts listed in table \ref{tab:allUncertainties}, we derive our final $\Gamma$-value:
\begin{equation}
\label{eqn:GammaFinal}
\Gamma=4376.210\;500\;872\;(102)(69).
\end{equation}
The first number in brackets is the statistical, the second one the systematic uncertainty.
\begin{table}
\centering
\caption{\label{tab:allUncertainties}Summary of the relative systematic shifts and uncertainties of $\Gamma$ ordered by the size of the relative uncertainties.}
\footnotesize
\begin{tabular}{c | l l }
\br
effect & rel. shift ($\times 10^{12}$) & rel. uncertainty ($\times 10^{12}$)\\
\mr
image charge shift & -282.4 & 14.1 \\
line-shape model of the dip & 0 & 5.2\\
magnetron frequency & 0 & 3.2 \\
dipole contribution of 2nd PnA pulse & 0 & $<3$\\
drift of the axial frequency & 0 & $\ll 1.2$ \\
motional magnetic field & 0 & $\ll 0.8$ \\
image current shift & -2.20 & 0.55 \\
electric field imperfections ($C_4$) & 0 & 0.50 \\
asymmetry  of the $\Gamma$-resonance & 0 & 0.3 \\
magnetic field imperfections ($B_2$) & 1.36 & 0.27 \\
fitting the $\Gamma$-resonance by a Gaussian & 0 & 0.1 \\
electric field imperfections ($C_6$) & 0 & 0.059\\
residual special relativity & -0.042 & 0.002\\
\mr
total relative shift & -283.3 & 15.4\\
\br
\end{tabular}\\
\end{table}
\normalsize
From this value, we derive the atomic mass of the electron according to equation (\ref{eqn:IndirectApproach}) using the theoretical \gf value for the bound electron. In the framework of bound-state QED the \gf of \CFive ~has been calculated with a relative uncertainty of $1.2\cdot10^{-11}:$ $g =2.001\;041\;590\;176\;(24)$~\cite{Strum2014_Emass}. 
Beyond that, the theoretical uncertainty can even be further reduced by a factor of 5 by estimating the dominant unknown term, the leading higher-order contribution in two-loop QED calculations: $g_{2L}^\tn{ho}(Z)\approx (\alpha/\pi)^2 (Z \alpha)^5 \; b_{50},$ from the discrepancy between the experimentally and determined \gf of hydrogen-like silicon, \SiThirteen, see supplement of~\cite{Strum2014_Emass}. We derive a final \gf value with a relative uncertainty of $2.3\cdot10^{-12}:$
\begin{eqnarray}
g_\tn{theo} &=2.001\;041\;590\;179\;8\;(47)\label{def:gFactorII}.
\end{eqnarray}
The required mass of the hydrogen-like carbon ion \CFive is calculated from the carbon atom as the unified atomic mass unit minus the five missing electrons and their binding energies as listed in table~\ref{tab:IonisationEnergies}.  
\begin{table}
\centering
\caption{\label{tab:IonisationEnergies}Binding energies of carbon, $^{12}\tN{C}$ from the \textit{NIST Atomic Spectra Database}~\cite{NIST_BindingEnergies}}
\footnotesize
\begin{tabular}{c l}
\br
ionization state  & binding energy (eV)\\
\mr
I~\cite{CI}   & $11.260\;30$\\
II~\cite{CII} & $24.384\;5\;(9)$\\ 
III~\cite{CIII}  & $47.887\;78\;(12)$\\ 
IV~\cite{CIV} & $64.493\;58\;(19)$\\ 
V~\cite{CV}  & $392.090\;49\;(3)$\\
\br
\end{tabular} 
\end{table}Since for the first ionization energy no uncertainty is listed in~\cite{CI} and a calculated value~\cite{CI_II} 
deviates $0.7 \di{meV}$, we assume a conservatively estimated error of $1\di{meV}$ for the first ionization state. With a total binding energy of
$E_\tn{bind}=540.116\;6\;(14)\di{eV}$ 
and the conversion factor of $1\di{u}=931.494\;061\;(21) \cdot 10^6\di{eV}$~\cite{CODATA2010}, the mass of \CFive ~is calculated:
\begin{eqnarray}
\label{def:MIon}
m(^{12}\tN{C}^{+5}) &= 12\di{u} -5 \cdot m_e+ E_\tn{bind}\nonumber\\
&= 11.997\;257\;680\;291\;7\;(18)\di{u} \tN{~~~~~~~~ (0.15 ppt)}.
\end{eqnarray}
With this we obtain the atomic mass of the electron\footnote{The value differs slightly ($0.15\sigma$) from the value in~\cite{Strum2014_Emass} due to a sign error in the image current shift.}:
\begin{equation}
\label{eqn:EmassFinal}
m_e=0.000\;548\;579\;909\;069\;4\;(128)(86)(13) \di{u}.
\end{equation}
The numbers in the brackets represent the statistical and systematic errors as well as the uncertainty of the theoretical \textit{g}-factor.\\
Our value is in agreement with the electron's atomic mass as listed in the 2010 CODATA tables of fundamental constants~\cite{CODATA2010} but has a 13 times higher precision. Particularly the phase-sensitive measurement technique, PnA, enabled a 19-fold improvement with respect to our previous most precise value of the electron's mass~\cite{Beier2003}.

\section{Conclusion}
In addition to the 13-fold improvement of the electron's atomic mass, our result has an impact on other fundamental constants or tests:\\ 
(1) The most precise value for the fine structure constant $\alpha$, which does not rely on QED in leading order, is derived from:
\begin{equation}
\label{eqn:alphaRecoil}
\alpha^2=\frac{2 R_\infty }{c}\frac{M_\tn{ref}}{m_e}\frac{h}{M_\tn{ref}}.
\end{equation}
It is based on measurements of the Rydberg constant $R_\infty$, the electron mass $m_e$ and the mass of an atom $M_\tn{ref}$, whose recoil velocity, $v_\tn{rec},$ is measured after the absorption of a single photon with the wavelength $\lambda$, to determine the ratio: $\frac{h}{M_\tn{ref}}=\lambda \cdot v_\tn{rec}.$ At present the precision in $\alpha$ is limited by the atom recoil experiment, measuring the ratio of the Planck constant and the rubidium mass with a relative uncertainty of $1.2\cdot 10^{-9}$ ~\cite{Bouchendria2011_Alpha}. In case of potential improvements of the $h/M_\tn{ref}$ determination, the present CODATA value for the electron mass would represent the largest error contribution. Our improved value for $m_e$ makes sure that this will not be the case ($\delta R_\infty/R_\infty=5\cdot 10^{-12}$ \cite{CODATA2010} and $\delta M_\tn{Rb}/M_\tn{Rb}=1.2\cdot 10^{-10}$ \cite{Myers_2010}).\\
(2) The search for possible limitations of the theory of quantum electrodynamics concentrates on strong electric or magnetic fields, where such limitations are most likely to occur. The electric field, seen by the electron at the Bohr radius of an hydrogen-like ion, provides the highest field strength in a bound system. The previous comparison of theory and experiment in \SiThirteen~\cite{Sturm2011_hydrogenlikeSi} was limited by the uncertainty of the electron's mass. Our new value allows for more significant tests of QED in strong electric fields by \gf measurements on hydrogen-like ions of higher nuclear charge values. Such experiments are currently assembled at HITRAP / GSI \cite{Kluge_2008} (ARTEMIS) and at MPIK (ALPHATRAP).\\
For an even more precise determination of the electron mass in the future a larger trap would significantly reduce the dominant systematic uncertainty of the image charge shift, which inversely scales with the third power of the trap radius. The statistical error, which scales with the width of the $\Gamma$-resonance, could be diminished, firstly by further cooling of the axial and cyclotron eigenmodes, reducing the thermal phase jitter, secondly by decreasing the phase detection uncertainty with either a very high-$Q$ resonator or a better compensated trap, enabling a larger amplification during the second PnA pulse.

\ack 
We thank our colleagues from theory Jacek Zatorski, Zolt\'an Harman and Christoph H. Keitel for stimulating discussions. We acknowledge financial support from the Max Planck Society, the EU (ERC grant number 290870; MEFUCO), the IMPRS-QD, GSI Helmholtzzentrum f\"ur Schwerionenforschung Darmstadt and the Helmholtz Alliance HA216/EMMI.

\section*{References}
\bibliography{iopart-num}

\end{document}